\documentclass[utf8,noinfo,a4paper]{frontiersSCNS_mod} 

\usepackage{url,lineno,microtype,subcaption}
\usepackage[colorlinks,citecolor=blue]{hyperref}

\newcommand{\eff}{\epsilon_{\mathrm{ff}}}
\newcommand{\tff}{t_{\mathrm{ff}}}
\newcommand{\msol}{\mbox{$\mathrm{M}_{\odot}$}}
\newcommand{\cs}{c_\mathrm{s}}
\newcommand{\lj}{\lambda_\mathrm{J}}
\newcommand{\mj}{M_\mathrm{J}}
\newcommand{\mach}{\mathcal{M}}
\newcommand{\macha}{\mathcal{M}_\mathrm{A}}
\newcommand{\cm}{\mathrm{cm}}
\newcommand{\pc}{\mathrm{pc}}

\newcommand{\Gauss}{\mathrm{G}}
\newcommand{\yr}{\mathrm{yr}}

\usepackage{color}
\definecolor{darkgreen}{RGB}{0,100,0}

\def\keyFont{\fontsize{8}{11}\helveticabold }
\def\firstAuthorLast{Krumholz \& Federrath} 
\def\Authors{Mark R. Krumholz\,$^{1,2,*}$ and Christoph Federrath\,$^{1,2}$}
\def\Address{$^{1}$Research School of Astronomy and Astrophysics, Australian National University, Canberra, ACT, Australia \\
$^{2}$Centre of Excellence for Astronomy in Three Dimensions (ASTRO-3D), Australia}
\def\corrAuthor{Mark R. Krumholz}

\def\corrEmail{mark.krumholz@anu.edu.au}

\begin{document}
\firstpage{1}

\title[Magnetic Fields, the SFR, and the IMF]{The Role of Magnetic Fields in Setting the Star Formation Rate and the Initial Mass Function} 

\author[\firstAuthorLast ]{\Authors} 
\address{} 
\correspondance{} 

\extraAuth{}

\maketitle

\begin{abstract}

\section{}
Star-forming gas clouds are strongly magnetized, and their ionization fractions are high enough to place them close to the regime of ideal magnetohydrodyamics on all but the smallest size scales. In this review we discuss the effects of magnetic fields on the star formation rate (SFR) in these clouds, and on the mass spectrum of the fragments that are the outcome of the star formation process, the stellar initial mass function (IMF). Current numerical results suggest that magnetic fields by themselves are minor players in setting either the SFR or the IMF, changing star formation rates and median stellar masses only by factors of $\sim 2-3$ compared to non-magnetized flows. However, the indirect effects of magnetic fields, via their interaction with star formation feedback in the form of jets, photoionization, radiative heating, and supernovae, could have significantly larger effects. We explore evidence for this possibility in current simulations, and suggest avenues for future exploration, both in simulations and observations.

\tiny
 \keyFont{ \section{Keywords:} galaxies: star formation, ISM: clouds, ISM: kinematics and dynamics, ISM: magnetic fields, MHD, stars: formation, turbulence } 
\end{abstract}

\section{Introduction}

Star-forming molecular clouds are threaded by magnetic fields that are likely inherited from the galactic-scale interstellar medium out of which they condensed (see the review by Hennebelle \& Inutsuka in this volume). These fields certainly influence cloud morphology and evolution. However, it remains an open question to what extent magnetic fields set the main quantitative outcomes of the star formation process: the rate at which molecular clouds convert their gaseous mass into stars, and the distribution of the masses of the resultant stars. The goal of this review is to summarize current observational and theoretical evidence that points toward a quantitative answer to these questions.

\subsection{Basic Physical Considerations}
\label{ssec:basic}

Any attempt to understand the role of magnetic fields in regulating the collapse and fragmentation of molecular clouds must begin from some basic physical considerations. The virial theorem provides a useful tool with which to describe the relative importance of magnetic forces in comparison to the forces of gravity, turbulent ram pressure, and thermal pressure. For a fixed control volume $V$ containing fluid of density $\rho$ and velocity $\mathbf{v}$, with magnetic field $\mathbf{B}$ and gravitational potential $\phi$, this is \citep{McKee92a}
\begin{eqnarray}
\lefteqn{\frac{1}{2}\ddot{I} = 2\left(\mathcal{T}-\mathcal{T}_0\right)
+ \left(\mathcal{B} - \mathcal{B}_0\right)
}
\nonumber \\
& &   {} + \mathcal{W} - \frac{1}{2} \frac{d}{dt}\int_{\partial V} \left(\rho\mathbf{v}r^2\right)\cdot d\mathbf{S},
\label{eq:vt}
\end{eqnarray}
where $\ddot{I}$ is the second derivative of the moment of inertia of the mass inside $V$, $\mathcal{T} = (1/2)\int (3P+\rho v^2)\,dV$ is the total translational thermal plus kinetic energy, $\mathcal{B} = (1/8\pi)\int B^2\,dV$ is the total magnetic energy (with $B \equiv |\mathbf{B}|$), $\mathcal{W}=-\int \rho \mathbf{r}\cdot\nabla\phi\,dV$ is the gravitational potential energy, and $\mathcal{T}_0$ and $\mathcal{B}_0$ represent the fluid and magnetic stresses, respectively, across the surface of $V$.  The right hand side of this equation expresses how the various forces together cause the material inside the volume to accelerate inward or outward. The final term, involving a time-derivative of the mass flux across the surface $\partial V$ of volume $V$, represents the change in inertia within the control volume not due to forces, but instead due to bulk flows of mass across the boundary.

Taking ratios of the force terms on the right-hand side of the virial theorem to form dimensionless ratios yields numbers that express their relative importance. Taking the ratio of the magnetic term to the two parts of the kinetic term yields
\begin{equation}
\frac{\mathcal{B}}{(3/2)\int P\, dV} \sim \frac{B^2/8\pi}{P} \sim \beta^{-1}
\end{equation}
and
\begin{equation}
\frac{\mathcal{B}}{(1/2)\int \rho v^2 \, dV} \sim \frac{B^2/8\pi}{\rho v^2} \sim \left(\frac{v_A}{v}\right)^2 \sim \mathcal{M}_A^{-2},
\end{equation}
where
\begin{equation}
v_{\rm A} = \frac{B}{\sqrt{4\pi\rho}}
\end{equation}
is the Alfv\'en speed. The quantities $\beta$ and $\macha$ are the plasma $\beta$ and Alfv\'en Mach number, respectively, and it is immediately clear that they describe the importance of magnetic forces in comparison to thermal and turbulent pressure. If $\beta \ll 1$, magnetic pressure greatly exceeds thermal pressure, and if $\mathcal{M}_A \ll 1$, magnetic pressure greatly exceeds turbulent pressure.

Similarly, taking the ratio of the magnetic and gravitational terms, and assuming that the volume's self-gravity dominates over any external field so that its gravitational energy may be expressed as $\mathcal{W} \sim -G M^2/R$, we have
\begin{equation}
\frac{\mathcal{B}}{\mathcal{W}} \sim \frac{B^2 R^3/8\pi}{GM^2/R},
\end{equation}
where $M$ is the mass within the volume and $R\sim V^{1/3}$ is its characteristic size. For ideal magnetohydrodynamics (MHD), the magnetic flux through the volume is fixed if there is no mass flux through its surface, and thus it is convenient to re-express this ratio in terms of the magnetic flux $\Phi_B \sim B R^2$, so that
\begin{equation}
\frac{\mathcal{B}}{\mathcal{W}} \sim \frac{\Phi_B^2}{G M^2} \sim \left(\frac{M_\Phi}{M}\right)^2 \sim \mu_{\Phi}^{-2},
\end{equation}
where
\begin{equation}
\label{eq:mphi}
M_\Phi \equiv \frac{1}{2\pi} \frac{\Phi_B}{\sqrt{G}}
\end{equation}
is the magnetic critical mass \citep{Mouschovias76a}, defined as the maximum mass that can be supported against collapse by a specified magnetic flux, and $\mu_{\Phi} = M/M_\Phi$ is the mass measured in units of $M_\Phi$. Clouds with $\mu_\Phi < 1$ are called magnetically subcritical, while those with $\mu_\Phi > 1$ are called magnetically supercritical. Note that the exact coefficient in $M_\Phi$ depends weakly on the configuration of the mass; the value $1/2\pi$ we have adopted in \autoref{eq:mphi} is for an infinite thin sheet \citep{Nakano78a}, but other plausible configurations give results that differ from this by only $\sim 10\%$ \citep{Tomisaka88a}.

Before moving on, we offer two cautions. First, the dimensionless ratios $\mathcal{M}_A$, $\beta$, and $\mu_{\Phi}$ that we have defined in order to characterize the importance of magnetic terms in the virial theorem do not include the surface fluid stress term $\mathcal{T}_0$, surface magnetic stress $\mathcal{B}_0$, and bulk flow term $(1/2)(d/dt)\int_V (\rho\mathbf{v}r^2)\cdot d\mathbf{S}$. Simulations show that these can make order unity contributions to the right hand side of \autoref{eq:vt} \citep{DibEtAl2007}, and the main reason we have omitted them is purely pragmatic: they are generally much more difficult to determine from observations than the volumetric terms. Nonetheless, we should keep in mind that conclusions about the relative importance of magnetic forces relative to others might be altered if we could properly include the hard-to-measure surface terms.

The second caution is that we have implicitly assumed that $\mu_\Phi$ is a constant, which is true only if the flux is conserved. This holds for ideal MHD, but non-ideal effects must become important at some point in the star formation process, as evidenced by the fact that the magnetic fields of young stars are far weaker than would be expected if all of the magnetic flux that threads a typical $\sim 1$ $M_\odot$ interstellar cloud were trapped in the star into which it collapses \citep[e.g.,][]{Paleologou83a}. Current simulations suggest that most loss of magnetic flux occurs on the scales of individual protostellar disks or smaller (e.g., \citealt{Tsukamoto15a, Tomida15a, NolanEtAl2017, Zhao18a, Wurster18a, Vaytet18a}; see \citealt{Li14a} for a review of earlier work), a scale that is mostly too small to be important for the SFR or the IMF. The non-ideal mechanism that operates on the largest scales is ion-neutral drift, also known as ambipolar diffusion, which allows a redistribution of magnetic flux in weakly-ionized plasma due to imperfect coupling between ions and neutrals. The importance of this mechanism can be characterized by the ambipolar diffusion Reynolds number $R_{\rm AD}$ \citep{Zweibel97a, LiEtAl2006, LiEtAl2008}, a quantity comparable to the classical fluid Reynolds number: the latter measures the ratio of the size scale of a turbulent flow to the size scale on which viscous dissipation occurs, while the former measures the ratio the flow size scale to the scale on which ions and neutrals are able to separate from one another. Observed dense molecular clumps have $R_{\rm AD}\approx 20$ \citep{McKee10a}, which places them close to but not strongly in the regime of ideal MHD (corresponding to $R_{\rm AD}\rightarrow\infty$). For this reason we will assume ideal MHD throughout most of this review, and briefly introduce non-ideal effects when they are particularly relevant.

\subsection{Historical and Observational Background}

Theories of how magnetic fields regulate the star formation rate (SFR) and the stellar initial mass function (IMF) can be classified is in terms of the assumptions they make, either implicitly or explicitly, about the values of the dimensionless ratios defined in \autoref{ssec:basic}. There is little doubt that $\beta < 1$, since molecular clouds are very cold and have low thermal pressures, but there is much more uncertainty about the values of $\mathcal{M}_A$ and $\mu_\Phi$. The dominant model of star formation prior to ca.~2000 implicitly assumed that molecular clouds also had both $\mathcal{M}_A < 1$ and $\mu_{\Phi} < 1$ \citep[e.g.,][]{Shu87a, Mouschovias99a}, i.e., their magnetic fields were strong enough that the pressure they provided was both stronger than the turbulent ram pressure and sufficient to prevent gravitational collapse. A model in which most molecular gas is subcritical leads to a picture of star formation in which the dominant physical processes are the non-ideal MHD mechanisms responsible for violation of flux-freezing, which allows $\mu_\Phi$ to increase until it is greater than unity (i.e., the cloud becomes supercritical) and collapse can proceed. This would imply that the rate of star formation is controlled by the rate at which mass is able to cross from $\mu_\Phi < 1$ to $\mu_\Phi > 1$ by non-ideal MHD effects \citep[e.g.,][]{TAssis04a, Shu07a}, and that the IMF is determined by the mass distribution of the resulting supercritical structures \citep[e.g.][]{Shu04a, Kunz09a}.

However, painstaking observational work in the past two decades, summarized in the review by \citet{Crutcher12a}, has called these assumptions into question. In particular, observations of Zeeman splitting provide a direct measurement of line-of-sight magnetic field strengths in molecular clouds, and Zeeman surveys have failed to detect a significant population of molecular clouds with $\mu_\Phi < 1$, in contrast to atomic clouds, which mostly have $\mu_\Phi < 1)$. For molecular gas they instead suggest a distribution of $\mu_\Phi$ values whereby $\mu_{\Phi}^{-1}$ is nearly flat from 0 to 1, i.e., clouds are uniformly distributed from nearly non-magnetized ($\mu_\Phi^{-1} = 0$) to lying on the boundary of super- and subcritical ($\mu_\Phi^{-1} = 1$). This would imply that the median molecular cloud has $\mu_{\Phi} \approx 2$, and is therefore supercritical. There are a few possible caveats to this conclusion. First, as noted above, a measurement of $\mu_\Phi$ only characterizes the importance of the volumetric magnetic field, not any potential contribution from magnetic stresses at cloud surfaces. Second, since the Zeeman effect only allows one to measure the line of sight magnetic field, inferences of the $\mu_{\Phi}$ distribution depend on statistical analysis of measurements along multiple sight lines under the assumption that magnetic field orientations along these sight lines are randomly distributed; if there are magnetic alignments over sufficiently large scales, this assumption might fail, in which case the statistical power of the conclusion would be reduced. Nonetheless, we regard these possibilities as unlikely, and so for most of this review we will adopt the view that observations favor $\mu_\Phi > 1$.

The value of $\macha$ is less certain. Observations of polarized thermal emission or polarized optical absorption by dust gains permit detection of the plane of the sky orientation of magnetic fields. These suggest that fields are relatively well-ordered \citep[e.g.][though in some cases alignment appears to break down at very small scales -- \citealt{Soam15a, Hull17a, Ching17a}]{HeyerBrunt2012, Li15b, Pattle17a, Soam18a}, and that they align well with structures in the gas column density \citep[e.g.][]{Planck-Collaboration16a}; simulations suggest that such features will be present only in flows with $\mathcal{M}_A \lesssim 1$  \citep[e.g.,][]{Li13b, Li15a, Federrath2016jpp, Tritsis16a, Tritsis18a, Mocz17a, TritsisEtAl2018}. On the other hand, \citet{Padoan99a} and \citet{Padoan04a} compare a wide range of statistics on the density, velocity, and magnetic field structure in molecular clouds to simulations with both $\macha \approx 1$ and $\macha \gg 1$, and conclude that only the latter are consistent with the observations. If $\macha \lesssim 1$, this would require that clouds be threaded by well-ordered fields with a significant net flux that dominate the total magnetic energy budget, while if $\macha \gtrsim 1$ the fields could be ordered, but they could also have a small net flux and be dominated by a disordered component \citep{Mac-Low99a,BruntFederrathPrice2010a}, such as that produced by a turbulent dynamo.

Regardless of whether $\macha \approx 1$ or $\macha \gg 1$, the observation that $\mu_\Phi > 1$ has led theoretical focus in the past few years to shift to models in which molecular clouds are assumed to be ``born'' supercritical \citep[e.g.,][]{Padoan99a, MacLowKlessen2004, Krumholz05a}, rather than having to transition to this state via some slow, non-ideal MHD process. In such a picture, the primary regulator of both the SFR and the IMF is usually assumed to be some combination of turbulence (strongly magnetized if $\macha \lesssim 1$, weakly magnetized otherwise) and stellar feedback; see \citet{Krumholz14a} for a recent review. In this context, magnetic fields are doubtless important for shaping the morphology of the ISM, particularly as regards to the filaments ubiquitously observed in both real molecular clouds and simulations. For example, magnetic fields clearly seem to play some role in determining the orientations of filaments \citep[e.g.,][]{Planck-Collaboration16a}, and may be responsible for setting their widths as well \citep[e.g.,][]{SeifriedWalch2015,Federrath2016,FederrathEtAl2016}. The relative orientations of magnetic fields and filaments appears to carry important information about whether flows in molecular clouds are predominantly solenoidal / shearing or compressive \citep{Soler17a}. However, it is not clear that these morphological factors are linked to the quantitative ``outputs'' of the star formation process, the SFR and IMF. Answering this question in the context of a cloud where $\mu_\Phi>1$ is the focus of the remainder of this review.

\section{Magnetic Fields and the Star Formation Rate}
\label{sec:sfr}

In this section we examine the question of how magnetic fields affect the rate of star formation in molecular clouds. We begin in \autoref{ssec:sfr_obs} with a brief review of the state of observations of the star formation rate, and in \autoref{ssec:sfr_models} we discuss recent theoretical and numerical work on the role that magnetic fields might play in explaining these observations. In \autoref{ssec:sfr_feedback} we highlight an important and but poorly explored frontier: the interaction between magnetic fields and stellar feedback.

\subsection{Observational Constraints on the Star Formation Rate}
\label{ssec:sfr_obs}

Star formation is a remarkably slow and inefficient process across nearly all size and mass scales. In nearby galaxies, the observed molecular gas depletion time (defined as the time required to convert all molecular gas to stars at the current star formation rate) at scales of $\gtrsim 100$ pc is $\sim 1$ Gyr \citep[e.g.][]{Bigiel08a, Blanc09a, Schruba11a, Rahman12a, Leroy13a, Leroy17a}. In comparison, the gas in molecular clouds has densities $n\gtrsim 30$ cm$^{-3}$, corresponding to free-fall times of at most $\tff = \sqrt{3\pi/32 G \mu_{\rm H} n} \lesssim 10$ Myr; here $\mu_{\rm H} = 2.34 \times 10^{-24}$ is the mean mass per H nucleus for standard interstellar medium (ISM) composition. This implies that the star formation rate is a factor of $\gtrsim 100$ smaller than what would be expected for clouds collapsing to stars in free-fall.

Formally, we can parameterize the efficiency of star formation in terms of the quantity $\eff$, defined such that a gas cloud of mass $M$, volume $V$, and free-fall time $\tff$ (evaluated at its mean density, $\rho = M/V$), and star formation rate $\dot{M}_*$ has
\begin{equation}
\eff = \frac{\dot{M}_*}{M/t_{\rm ff}}.
\end{equation}
Intuitively $\epsilon_{\rm ff}$ represents the ratio of the observed star formation rate in a region to the maximal rate that would be expected if gas were to collapse in free-fall with nothing to inhibit it. Normalizing to $t_{\rm ff}$ is critical when one wishes to compare samples across a wide range of size and density scales, since denser objects invariably have higher star formation rates per unit mass simply as a result of their shorter dynamical times. If one does not remove the dependence on dynamical time by measuring $\eff$, rather than, for example, the specific star formation rate $\dot{M}_*/M$, then anything that correlates with density will appear to correlate with star formation activity.

\subsubsection{Counts of Young Stellar Objects}

The observations discussed above imply that, measured at kpc scales, $\eff \lesssim 0.01$. However, it is possible to constrain $\eff$ more precisely, and on smaller scales, with a variety of techniques. The most direct method is simply to count young stellar objects (YSO) within resolved nearby molecular clouds. If one knows the mean duration of the observed YSO phase (e.g., if the observed YSOs are selected based on the presence of 24 $\mu$m excess, which several lines of evidence suggest persists for $\approx 2$ Myr -- \citealt{Evans09a}), then the mass of YSOs in that phase provides an estimate of the star formation rate. Combining this with a measurement of a mass and an estimate of the volume density (uncertain since the line of sight depth of a cloud cannot usually be measured directly), yields an observational estimate of $\eff$. In the past decade a number of studies have been published using this methodology \citep{Krumholz12a, Lada13a, Federrath2013sflaw, Krumholz14a, Evans14a, Salim15a, Heyer16a, Ochsendorf17a}, and all published studies are consistent with an estimate $\eff \approx 0.01$, with roughly a factor of 3 scatter and a factor of 3 systematic uncertainty, mainly coming from uncertainties in the gas density and the duration of the observed YSO phase.\footnote{Note that \citet{Ochsendorf17a} measure $\eff$ in molecular clouds in the Large Magellanic Cloud using two separate methods: counts of massive ($M \gtrsim 8$ $M_\odot$) YSOs, and a cloud matching technique as described below. Our statement here applies to their YSO counting method, which gives a distribution of $\eff$ with a median of $\log \eff = -1.7$ and a 16th to 84th percentile range from $\log \eff=-2.03$ to $-1.25$, consistent with both the median and the spread of the other YSO counting studies within the systematic uncertainty. By contrast their cloud matching method gives a median $\log \eff=-1.3$ with a 16th to 84th percentile range $\log \eff = 1.74$ to $-0.69$, as we discuss below. The numerical median and percentile ranges we quote are compiled by \citet{Krumholz18b}, who derive them from Table 6 of \citet{Ochsendorf17a}.}

There is at present no evidence for systematic variation of $\eff$, as opposed to systematic variation in the overall or specific star formation rate, with properties of the magnetic field. To date the only published study searching for magnetic effects on the star formation rate from observation is that of \citet{Li17a}, who analyze the cloud samples of \citet{Lada10a} and \citet{Heiderman10a}. They define the orientation of a cloud on the sky as the direction in which the observed extinction map has the largest autocorrelation, and find that the star formation rate per unit mass is systematically higher in clouds where the large-scale magnetic field and cloud orientation vectors are closer to parallel. However, \citet{Krumholz12a} analyzed the same samples and found that $\eff$ is nearly the same in all of the clouds they contain. Consequently, the most natural interpretation of the \citet{Li17a} study is not that magnetic fields have an important effect on the star formation rate, but instead that denser clouds are more likely to have magnetic fields oriented along rather than orthogonal to their long axis, and that the apparent correlation between star formation and magnetic fields is simply a result of both correlating with density.  In order to demonstrate that magnetic fields (or any other cloud property) is changing the nature of the star formation process, one would need to show not merely that the star formation rate as a whole changes with that property, but that the star formation rate per dynamical time (i.e., $\eff$) does. There is some evidence for such variations in $\eff$ as a function of Mach number \citep[e.g.,][]{Federrath2013sflaw,Salim15a,ShardaEtAl2018}, but there have been no comparable observational efforts to search for  variations in $\eff$ as a function of magnetic properties.

\subsubsection{Alternative Methods}

While YSO counting is the most direct and unambiguous method of estimating $\eff$, one can only use it in relatively nearby clouds due to the need to resolve individual YSOs.\footnote{As noted above, it is possible to extend the YSO counting method to the Magellanic Clouds, but at the price of substantially reduced sensitivity and increased uncertainty, because at such large distances observations can at present detect only very massive YSOs, $M \gtrsim 8$ $M_\odot$ \citep{Ochsendorf16a, Ochsendorf17a}, which must then be extrapolated to estimate the mass of the unseen population of lower mass stars. Both this extrapolation and timescales of massive YSO evolution (needed to complete the estimate of $\eff$) are substantially uncertain.} More distant targets require different methods. Three in common use are pixel statistics, the HCN to IR ratio, and cloud matching. The method of pixel statistics is to map the distributions of molecular gas and star formation in an external galaxy at high spatial resolution -- typically tens of pc for the gas. The molecular gas map provides both the gas surface density and the velocity dispersion; the latter, together with an estimate of the stellar surface density, allows one to estimate the midplane volume density from hydrostatic equilibrium. Thus in each pixel one has available mass, free-fall time, and star formation rate, yielding an estimate of $\eff$. Studies using this method thus far yield $\eff$ with a dispersion comparable to that produced by YSO counting, but with a factor of $\sim 2-3$ lower mean \citep{Leroy17a, Utomo18a}; given the systematic uncertainties in the methods, this is consistent with the distributions of $\eff$ being the same.\footnote{Of course we cannot definitively rule out the possibility that there is in fact a systematic difference between the Milky Way and the LMC (the only two systems for which YSO counting is available) and the slightly more distant galaxies surveyed by \citet{Leroy17a} and \citet{Utomo18a}. However, systematics due to the differences in method seem the more likely explanation.} The HCN method exploits the fact that, because it is subthermally-excited at low density, HCN traces ISM at densities $\gtrsim 10^4$ cm$^{-3}$ \citep{Shirley15a, Onus18a}, and thus one can estimate the local gas density producing HCN emission even if the emitting region is unresolved.\footnote{\citet{Kauffmann17a} argue that the density traced by HCN can be a factor of a few smaller if molecular clouds host a significant free electron population, which would help excite the HCN at lower densities. It is unclear at present to what extent \citeauthor{Kauffmann17a}'s result, which is derived based on high-resolution observations of a single nearby source, can be extrapolated to the much larger scales on which HCN is generally used as a diagnostic of $\eff$.} If one also uses a radiative transfer calculation to estimate the HCN emitting mass and correlates this with a tracer of the star formation rate (most commonly infrared luminosity), this provides all the ingredients necessary -- mass, star formation rate, and free-fall time -- to constrain $\eff$. As with pixel statistics and YSO counting, the result of this procedure is generally that $\eff \approx 0.01$ with a factor of $\sim 3$ dispersion and a comparable systematic uncertainty \citep[e.g.][]{Wu10a, Usero15a, Stephens16a, Onus18a, Gallagher18a}.

In the cloud matching technique, one constructs catalogs of molecular clouds and star-forming regions, and matches them up based on criteria of separation in position and velocity space. For each pair of matched clouds and star-forming regions, one infers the star formation rate of the star-forming region from its luminosity in IR or radio, and the mass and free-fall time of the cloud from its molecular line emission, yielding an estimate of $\eff$. In contrast to all other methods, for which the distribution of $\eff$ values inferred generally has a dispersion of only $\lesssim 0.5$ dex, cloud matching yields much larger dispersions of $\gtrsim 0.8$ dex, with some surveys producing a tail of clouds with $\eff \approx 1$ \citep{Vutisalchavakul16a, Lee16a, Ochsendorf17a}. In some of these studies the mean value of $\eff$ is also substantially higher than the value of $\eff \approx 0.01$ found by other methods. The difference in results cannot simply be a result of the cloud matching surveys targeting different regions or types of molecular cloud than the other studies, in part because cloud matching studies of the same region are often inconsistent with one another -- \citet{Vutisalchavakul16a} and \citet{Lee16a} both studied the inner Milky Way, but obtained median values of $\eff$ that differ by $\approx 0.8$ dex. 

Instead, the source of the discrepancy between the different cloud matching studies, and between cloud matching and other methods, appears to be in the process of constructing the cloud and star-forming region catalogs and matching them to one another. Both molecular gas emission and star formation tracer maps are continuous or nearly so, particularly toward molecule-rich regions such as the inner Milky Way. The process of breaking these continuous maps up into discrete ``clouds'' and ``star-forming complexes'' necessarily involves choices about how to perform the decomposition, and because the ``clouds'' and ``complexes'' are not co-spatial, these choices must be made independently for each map, and then one must decide how to associate the ``clouds'' in one map with the ``complexes'' in the other. Depending on how one makes these choices, a wide range of outcomes are possible. The difference between the results of \citet{Vutisalchavakul16a} and \citet{Lee16a} arise primarily from the fact that \citeauthor{Vutisalchavakul16a} use substantially more restrictive criteria for matching clouds with H~\textsc{ii} regions, and decline to estimate $\eff$ values for H~\textsc{ii} regions for which they cannot confidently identify a parent cloud. \citeauthor{Lee16a} are much less restrictive in their matching. This problem is unique to cloud matching, because in all the other techniques (YSO counting, pixel statistics, and HCN) the star-forming tracer and the molecular gas are co-spatial, so however one chooses to break up maps of one, it is possible to use the same decomposition for the other.

Given this review of the observational literature, our tentative summary is that observations require that $\eff \approx 0.01$ appears to be ubiquitous across spatial scales, from kpc-sized swathes of galaxies to individual molecular clouds and clumps $\approx 1$ pc in size, at densities up to $\sim 10^4$ cm$^{-3}$. This leads us to the central question for \autoref{sec:sfr}: to what extent can magnetic fields in supercritical molecular clouds help explain this observation?

\subsection{Magnetic Regulation of the SFR in Supercritical Clouds}
\label{ssec:sfr_models}

In a cloud that is magnetically supercritical, magnetic fields alone cannot significantly inhibit collapse. To see this, one need merely examine the magnetic and gravitational terms in the virial theorem (\autoref{eq:vt}). For a cloud of mass $M$ and radius $R$ threaded by a uniform magnetic field $B$, the gravitational and magnetic terms in the virial theorem can be expressed as $\mathcal{W} \sim G M^2/R$ and $\mathcal{B} \sim G M_\Phi^2/R$, respectively; recall that $M_\Phi$ is the maximum mass that can be supported by the magnetic field. The key point to notice is that both these terms scale with radius as $1/R$, so that even if $|\mathcal{W}|$ is only slightly larger than $\mathcal{B}$ when a cloud is at some starting characteristic size $R_0$, the mismatch between these two terms will grow as the cloud contracts, such that, by the time the cloud has been reduced to a size $\sim R_0/2$, $|\mathcal{W}|$ will be larger than $\mathcal{B}$ by a factor of 2, and the collapse will accelerate only a factor of 2 slower than if the magnetic field were absent entirely. The point to take from this thought exercise is that, due to the $1/R$ scalings of the gravitational and magnetic terms in the virial theorem, even a magnetic field that nearly strong enough to render a cloud subcritical at the start of its life will only slightly delay collapse.\footnote{Our claim that $\mathcal{B}$ will become increasingly unimportant compared to $\mathcal{W}$ as a cloud collapses might fail if the collapse drives a significant dynamo. In this case the dynamo would cause an increase in the magnetic energy $\mathcal{B}$ without a concomitant increase in the net magnetic flux, so that our assumption that $\mathcal{B}\propto M_\Phi^2$ would fail \citep{Birnboim18a}. However, even if this does occur, since the dynamo is ultimately powered by the collapse, it is energetically limited to $\mathcal{B} < f |\mathcal{W}|$ for some $f < 1$. Thus our claim that a magnetic field can only delay collapse in a supercritical cloud by a factor of order unity continues to hold.} To the extent that ion-neutral drift is important, it only strengthens this conclusion, since this mechanism tends to decrease the magnetic flux and thus $M_\Phi$ in the densest regions. For this reason, we focus on the role of magnetized turbulence in regulating star formation rates, rather than on magnetic fields by themselves.

\subsubsection{Star Formation Rates from Magnetized and Non-Magnetized Turbulence}

\begin{figure*}
\begin{center}
\includegraphics[width=\textwidth]{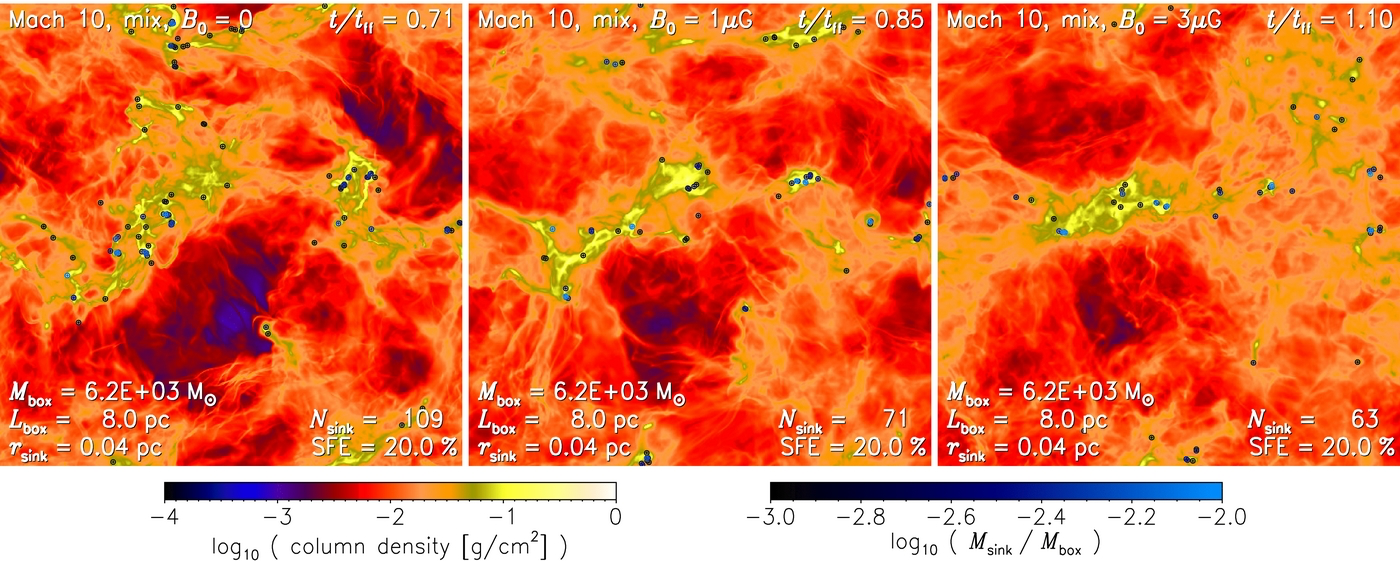}
\end{center}
\caption{
\label{fig:mhd_turb}
Density projections in three simulations of self-gravitating MHD turbulence from \citet{Federrath12a}. Each simulation takes place in a periodic box $8$ pc on a side, initially containing $6200$ $M_\odot$ of isothermal gas with sound speed $c_s = 0.2$ km s$^{-1}$, driven with a mix of solenoidal and compressive turbulent modes at a sonic Mach number $\mathcal{M} = 10$. The three simulations were initialized with uniform magnetic fields with field strength $B=0$, $1$ $\mu$G, and $3$ $\mu$G (left to right); once the turbulence reaches steady state, the corresponding Alfv\'en Mach numbers are $\macha = \infty$, $13$, and $2.7$. Points show the locations of sink particles, with color indicating mass. All three simulations have been run to the point where 20\% of the initial gas mass has converted to stars, but they have taken different lengths of time to reach this point.
}
\end{figure*}

What do simulations tell us about the star formation rate of magnetized versus unmagnetized turbulence? Here we focus on this question in the context of pure turbulence, deferring the question of the interaction of magnetic fields with stellar feedback to \autoref{ssec:sfr_feedback}. We show an example result from numerical simulations in \autoref{fig:mhd_turb}. As the figure shows, the presence of a dynamically significant magnetic field generally reduces the density contrast in turbulence, and leads to a pattern of star formation that is more distributed. The overall star formation rate decreases, or equivalently the time required to convert a fixed fraction of the gas to stars increases, as the magnetic field strength increases. A number of authors have conducted parameter studies of the star formation rate in supersonic turbulence as a function of magnetic field strength and other parameters \citep{PriceBate2009, Dib10a, Padoan11a, Padoan12a, Federrath12a}. The primary finding from these studies is that, compared to a non-magnetized flow of equal Mach number and virial ratio (ratio of kinetic to gravitational potential energy), a magnetic field strong enough to render the gas trans-Alfv\'enic ($\macha \approx 1$) but still leave it supercritical ($\mu_\Phi < 1$) results in a star formation rate that is a factor of $\approx 2-3$ lower. This finding holds over a range of sonic Mach numbers $\mathcal{M} \approx 5 - 50$ and cloud virial ratios $\alpha_{\rm vir} \approx 1 - 5$. These findings indicate that magnetic fields by themselves cannot explain the low value of $\eff$, but that they can contribute non-negligibly towards an explanation.

The mechanism by which magnetic fields reduce the star formation rate is not entirely clear. Modern theories that attempt to explain the low value of $\eff$ as a consequence of turbulence generally contain a few basic ingredients \citep[e.g.,][]{Krumholz05a, Padoan11a, Padoan12a, Hennebelle11a, HennebelleChabrier2013, Federrath12a, Hopkins12a, Hopkins13a, Burkhart18a, Burkhart18b}. The first of these is that turbulence, possibly coupled with self-gravity, will impose a certain probability distribution function (PDF) on the gas density. In the simplest models this PDF is taken to be log normal, since numerous numerical and analytic studies show that isothermal, non-self-gravitating turbulence generates a PDF of this form. However, some models also add a time-dependent evolution of the high-density tail into a power law shape, since simulations of turbulence with self-gravity show that such tails tend to grow over time \citep[e.g.,][]{Klessen2000, DibBurkert2005, Kritsuk11a, Collins11a, Collins12a, FederrathKlessen2013, GirichidisEtAl2014, Burkhart17a, Scannapieco18a}. The second is that the presence of turbulent motions imposes a critical density at which molecular clouds transition from gravitationally unbound and inert to bound and star-forming. Depending on the model, this density may be uniform everywhere, or it may depend on the particular length or size scale. Third, mass within a molecular cloud that exceeds the density threshold for stability is assumed to collapse into stars and be replaced with fresh, lower density material on some timescale. Again, depending on the model this timescale can be the local free-fall time in the high-density gas, the mean-density free-fall time of the entire cloud, or anything in between.

Models based on this paradigm of turbulent regulation appear to be able to reproduce a broad range of observables. For example, \citet{Padoan17a} simulate a large section of a galaxy in which molecular cloud turbulence is driven by supernovae; they study the distribution of $\eff$ values within individual molecular clouds, and find a median value of about $0.025$, with a spread of $\approx 0.5$ dex, fully consistent with the observed distribution. Similarly, \citet{Semenov16a} use a turbulence-regulated star formation prescription as a subgrid model in a galaxy-scale simulation, and show that the result agrees well with galactic-scale measurements of the correlation between star formation and gas surface densities.

In the context of these models, magnetic fields play a few potentially important roles, which in general tend to lower the star formation rate compared to a similar non-magnetized case. First, the presence of a magnetic field narrows the density PDF compared to what would prevail in a non-magnetic flow, because magnetic fields provide an additional support against shock compression that renders it more difficult to drive gas to high densities. This narrowing will lead to less mass exceeding the threshold density for the onset of collapse. This effect has been studied by a number of authors \citep[e.g.,][]{Cho03a, Kowal07a, Burkhart09a, Molina12a, Mocz17a}, but its magnitude is still not entirely certain, because it depends crucially on the scaling of magnetic field strength with density. The density jump across an isothermal shock of sonic Mach number $\mathcal{M}$ with pre-shock ratio of thermal to magnetic pressure $\beta_0$ will depend on how the pre- and post-shock magnetic fields compare, which is determined by the relative orientation between the field and the shock plane. This distribution of relative orientations is most conveniently expressed in terms of the magnetic field-density scaling.

For a constant magnetic field on both sides of the shock, expected if the typical shock is orthogonal to the local magnetic field, the density jump is $\rho_1/\rho_0 \propto \mathcal{M}^2$ independent of $\beta_0$, while for $B\propto \rho^{1/2}$, for example, $\rho_1/\rho_0 \propto \mathcal{M}^2 \beta_0/(\beta_0+1)$; more detailed expressions for other scalings may be found in \citet{Molina12a} and \citet{Mocz18a}. In the regime of super-Alfv\'enic turbulence ($\macha \gg 1$) and in the absence of self-gravity, the turbulence is isotropic and both analytic arguments and simulations predict the latter scaling, $B\propto \rho^{1/2}$ \citep[e.g.][]{Collins11a, Collins12a}. This leads to a prediction that the variance of the logarithmic density distribution depends on mean Mach number and plasma $\beta$ as \citep{Molina12a}
\begin{equation}
\label{eq:sigma_lnrho}
\sigma_{\ln \rho}^2 = \ln \left(1 + b^2\mathcal{M}^2 \frac{\beta}{\beta+1}\right),
\end{equation}
where $b$ is a constant of order unity that depends on the turbulent driving pattern \citep{FederrathKLessenSchmidt2008, KonstandinEtAl2012ApJ, FederrathBanerjee2015}. When $\beta \ll 1$, as is the case for observed molecular clouds,\footnote{For 10 K gas that is 75\% H$_2$ and 25\% He by mass, typical properties in a molecular cloud, $\beta = 0.21 n_3/B_1^2$, where $n_3$ is the number density of H nuclei in units of $10^3$ cm$^{-3}$ and $B_1$ is the magnetic field strength in units of 10 $\mu$G. \citet{Crutcher12a} finds typical field strengths $B_1\approx 1$ at $n_3 \approx 1$, corresponding to $\beta \sim 0.1$, and $B_1 \approx 500$ at $n_3 \approx 1000$, corresponding to $\beta\sim 10^{-3}$.} this yields a significantly lower dispersion of densities than for a non-magnetized flow, $\beta = \infty$.

However, this relation breaks down in the trans- or sub-Alfv\'enic regime that we have argued above is likely more realistic. For such flows, the magnetic field appears to suppress the density variance less than what would be predicted by \autoref{eq:sigma_lnrho}. This may be because the anisotropy of sub-Alfv\'enic turbulence means that one can no longer assume a single, simple density-magnetic field scaling. For example, if strong magnetic fields confine turbulent motions to flow primarily along rather than across field lines, then most shocks will be predominantly orthogonal to the field, in which case the pre- and post-shock fields will be nearly identical, and magnetic forces will not provide any resistance to compression. It is also unclear if the scaling between $B$ and $\rho$ might be different for strongly self-gravitating flows. \citet{Li15a} find in simulations of the formation of an infrared dark cloud that volume-averaged density and magnetic field strengths are related by $\langle B\rangle \propto \langle \rho \rangle^{0.65}$, but it is unclear if the same powerlaw relationship applies point-wise, rather than averaged over volumes. In their self-gravitating simulations, \citet{Mocz17a} find scalings that vary from $B\propto \rho^{2/3}$ for initially-weak fields ($\macha \gg 1$) to $B \propto \rho^{1/2}$ for initially-strong fields, with a smooth transition as a function of $\macha$. In order to fully understand how magnetic fields modify the density PDF, more studies of this type, across a wider range of parameter space, will be needed to extend the \citet{Molina12a} scaling. In addition, there is a need for more extensive studies including the effects of ion-neutral drift. Only a few studies of this type have been published \citep{LiEtAl2008, Downes2012, Meyer14a, Burkhart15a, Ntormousi16a}, and they suggest that ion-neutral drift at the levels expected for molecular clumps with the observed value $R_{\rm AD}\sim 20$ should partially offset the tendency of magnetic fields to narrow the density PDF, increasing the width back toward that produced in the non-magnetized limit. However, there has yet to be a comprehensive survey of parameter space.

A second way that magnetic fields can alter the star formation rate is by providing additional support against collapse, and thereby increasing the density threshold at which self-gravity becomes dominant. Consider a uniform spherical region of radius $R$, density $\rho$, 1D velocity dispersion $\sigma$, and magnetic field $B$; for this region, the condition for the right-hand side of the virial theorem (\autoref{eq:vt}) to be negative and thus indicative of collapse is, neglecting the surface terms
\begin{eqnarray}
\rho & > & \frac{3}{4\pi G R^2}\left(c_s^2 + \sigma^2 + \frac{v_A^2}{2}\right)
\nonumber \\
& = & \frac{3}{4\pi G R^2}\left[\left(1+\beta^{-1}\right)c_s^2 + \sigma^2\right].
\label{eq:rho_crit}
\end{eqnarray}
Thus a non-zero magnetic field, implying $v_A > 0$, makes it more difficult for a small-scale structure to collapse. A number of authors have suggested modified collapse criteria incorporating effects similar in functional form to \autoref{eq:rho_crit} \citep{HennebelleChabrier2008, HennebelleChabrier2009, Padoan11a, Federrath12a, Hopkins12a, Hopkins13a}. However, we caution that none of these modifications (nor, indeed, their original unmagnetized versions) properly account for the surface terms in the virial theorem, which can be non-negligible \citep{DibEtAl2007}.

As with the density PDF, the importance of this effect depends on the small-scale magnetic field and its correlation with density: if $B \propto \rho^{1/2}$, as expected for super-Alfv\'enic, non-self-gravitating flows, this would imply $v_A \approx \mbox{constant}$, in which case magnetic effects would impose a very important modification on the collapse criterion, because in observed molecular clouds $v_A / c_s \gtrsim 10$, so a magnetic field would have the effect of raising the effective sound speed of the gas by a factor of a few to ten. However, this may be an overestimate of the true effect, because the $B \propto \rho^{1/2}$ scaling follows only on scales where the turbulence is super-Alfv\'enic. Dense regions in turbulent media have smaller velocity dispersions, both because they tend to be physically small, and because density and velocity are anti-correlated \citep[e.g.,][]{Offner09a}, and thus at scales dense enough to be candidates for collapse the $B\propto \rho^{1/2}$ scaling might break down because the field is anisotropic. \citet{Hopkins13a} suggest an alternate collapse criterion that attempts to take this effect into account, but thus far it has not been tested in simulations.

Given the uncertainty on the scaling of magnetic field with density, it is not entirely clear which of the two mechanisms we have discussed -- narrowing of the density PDF or increasing the threshold for collapse -- is dominant in explaining how magnetic fields lower the star formation rate, or if both contribute comparably. Although they have not been explored extensively, for completeness we mention two other possible mechanisms that seem worth of investigation. First, one crucial ingredient of turbulence regulation models is the velocity power spectrum, which determines the scaling between $\sigma$ and $R$ in \autoref{eq:rho_crit} and analogous collapse conditions. There is limited evidence from some MHD simulations that the presence of a strong magnetic field might alter the velocity power spectrum \citep[e.g.,][]{Lemaster09a, Collins12a}, but the issue has received only limited exploration, and all published analytic models to date assume the same velocity power spectrum for magnetized and non-magnetized flows. Thus the potential impact of a velocity power spectrum that depends explicitly on magnetic field strength has not been explored. A second potential effect of magnetic fields is in models that include a powerlaw tail in the density PDF. The rate at which such tails develop, and the density at which they join onto the log normal part of the PDF, are at least potentially sensitive to the magnetic field strength. At present, however, no published models have examined this possibility. However, we emphasize that, while the \textit{mechanism} by which magnetic fields reduce the star formation rate in a turbulent medium relative to the non-magnetized case is uncertain, the numerical experiments leave little doubt that the \textit{amount} of reduction is roughly a factor of two to three, at least in the ideal MHD limit.

\subsubsection{Effects on Maintenance of Turbulence}

In addition to directly reducing the rate of star formation via their effects on the gas density structure and boundedness, magnetic fields may also affect the star formation rate in turbulent flows in two other ways. The first, via their effect on the rate at which turbulence decays, we discuss here, while the second, through their interaction with feedback, we defer to \autoref{ssec:sfr_feedback}.

\begin{figure*}
\begin{center}
\includegraphics[width=8cm]{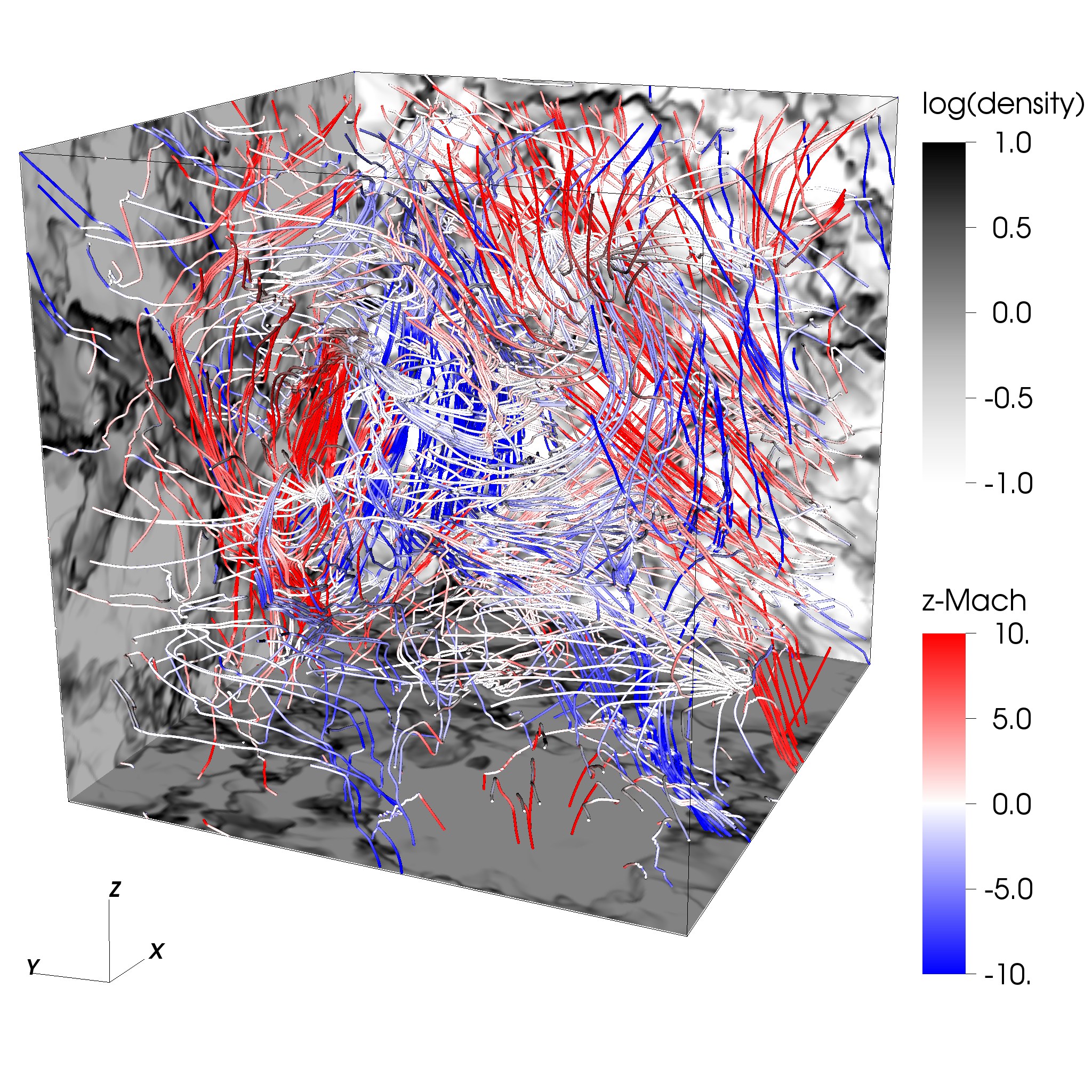}\includegraphics[width=8cm]{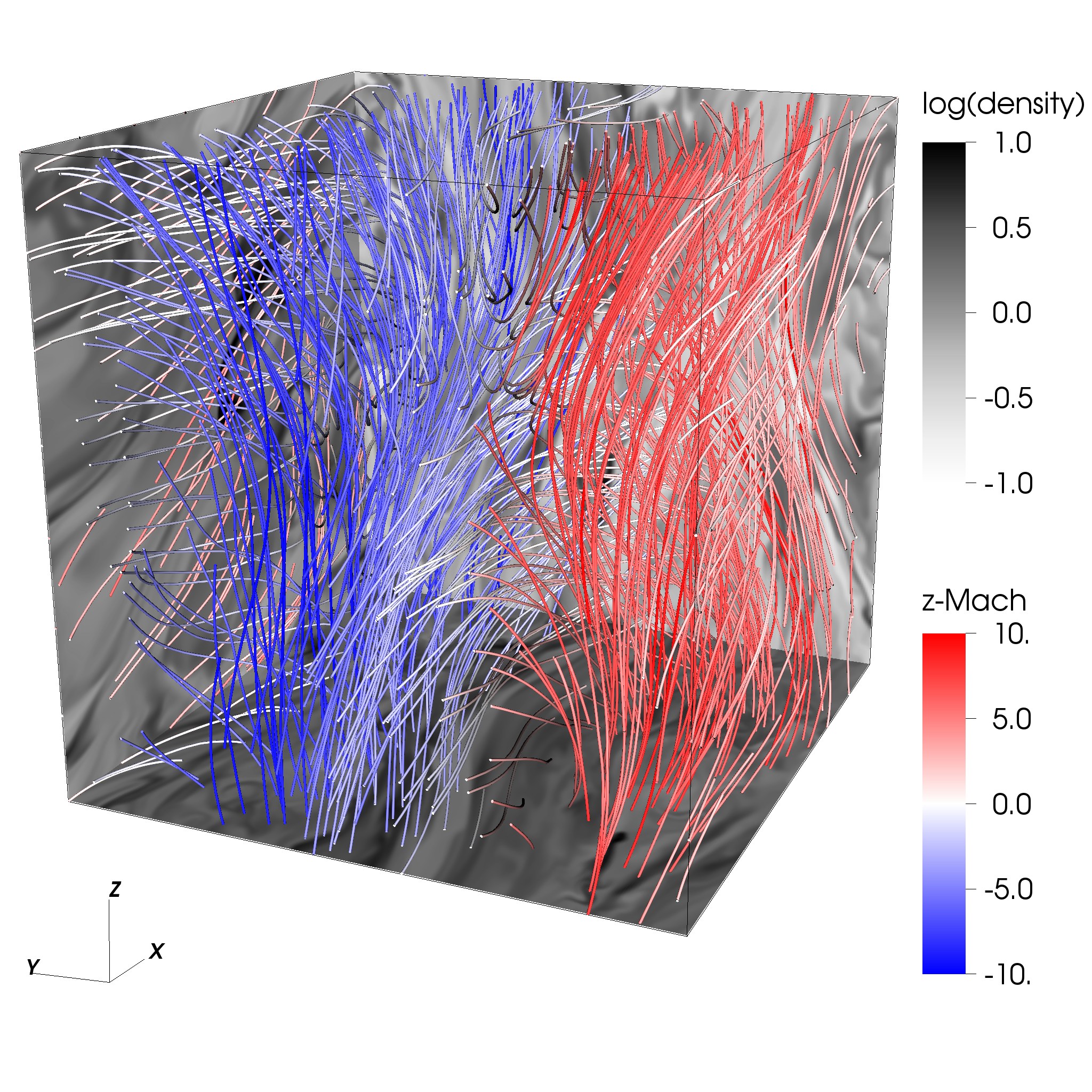}
\end{center}
\caption{
\label{fig:birnboim18}
Results from two simulations of compressing isothermal turbulence, without (\textbf{left}) and with (\textbf{right}) a magnetic field, from \citet{Birnboim18a}. In both panels, lines show flow streamlines, with the $z$-velocity along the streamline color-coded by Mach number. Grayscale on the box edges indicates the logarithm of gas density, in units where the mean density in the computational domain is unity. The total velocity dispersions in the two simulations shown are comparable, but the simulation including a magnetic field has a much lower dissipation rate because the field has organized the flow into a highly-anisotropic state.
}
\end{figure*}

One of the fundamental challenges in understanding the low observed value of $\eff$ via turbulence is that supersonic turbulence decays on a time scale comparable to the turbulent flow crossing time, which, in a system with virial ratio near unity, is comparable to the free-fall time \citep[e.g.,][]{Tan06a}. By itself, the presence of a magnetic field does not appear to change this basic result \citep[e.g.,][]{Stone98a, Mac-Low98a, Mac-Low99a, Ostriker99a, HeitschMacLowKlessen2001}; at best strongly magnetized thin sheets can retain a small amount of excess kinetic energy in the form of incompressible motions in the sheet \citep{Kim13d}. However, there is one possible exception to this statement worth noting: while magnetic fields do not alter the decay rate of turbulence driven by external forcing, for example by star formation feedback, it is possible that they do alter the decay rate of turbulence that is driven by the self-gravitational compression of the gas itself \citep{Birnboim18a}. This effect is driven mainly because compression in a strongly magnetized gas causes the flow to become highly anisotropic, and anisotropy reduces the decay rate of supersonic turbulence because the decay rate becomes of order the crossing time in the most elongated, slowest direction \citep{Cho03a, Hansen11a}. We illustrate this effect in \autoref{fig:birnboim18}. Consequently, while a compressing hydrodynamic fluid will remain turbulent only as long as the compression timescale is comparable to or smaller than the crossing timescale \citep{Robertson12a}, for a magnetized compressing fluid this requirement is considerably relaxed.

This effect has yet to be embedded in the context of an analytic or semi-analytic model, and simulations of collapsing magnetized clouds have generally included other physical mechanisms, particularly star formation feedback or thermal instability, that would make it hard to isolate the importance of this effect. Nonetheless, it seems possible that the increased efficiency of turbulent driving in a magnetized compressing medium relative to a non-compressing one may be important for explaining the ubiquity of turbulent motions observed in molecular clouds and the low value of $\eff$ that they appear to produce.

\subsection{Magnetic Fields and Feedback}
\label{ssec:sfr_feedback}

Perhaps the most important possible effect of magnetic fields on star formation rates is via their interaction with feedback. A full review of feedback mechanisms is beyond the scope of this paper, and we refer readers to \citet{Krumholz14b}. Here we focus on the interaction of feedback mechanisms with magnetic fields, and the impact of this interaction on star formation rates.

\subsubsection{Protostellar Outflows}

As mass falls onto forming stars, its angular momentum causes it to form disks, and matter orbiting in disks creates helical magnetic fields that launch some fraction of the accreting material into a fast-moving outflow \citep[and references therein]{Bally16a}. Magnetic fields (and possibly also non-ideal MHD effects -- e.g., \citealt{Tomida15a, Tsukamoto15a, NolanEtAl2017, Zhao18a}) are clearly required for launching outflows in the first place. However, they also play a crucial role in regulating their interaction with the surrounding environment. Protostellar outflows are highly-collimated: \citet{Matzner99a} show that, far from their launch point, all hydromagnetic winds approach a common momentum distribution
\begin{equation}
\frac{dp}{d\mu} \propto \frac{1}{\ln\left(2/\theta_0\right)\left(1+\theta_0^2-\mu^2\right)},
\end{equation}
where $\mu = \cos\theta$, $\theta$ is the angle relative to the central axis of the outflow, and $dp/d\mu$ is the differential momentum carried by the wind within a range of angles $\mu$ to $\mu+d\mu$. The parameter $\theta_0$ specifies the intrinsic breadth of the outflow, and is typically small, implying a high degree of collimation: \citet{Matzner99a} estimate $\theta_0 \approx 0.01$, which corresponds to 50\% of the total outflow momentum being injected into 1\% of the solid angle centered on the outflow axis.

Due to this high degree of collimation, for purely hydrodynamic flows (even if we neglect the fact that without magnetic fields no outflows would form at all), the effects of outflows should be very limited. Since pressure forces are generally negligible in molecular clouds, there is no efficient mechanism to redistribute the narrowly-focused outflow momentum. Consequently, one excepts that outflows will simply punch small holes into their parent clouds. Magnetic fields, on the other hand, couple gas across larger distances, and thus do provide a mechanism by which the momentum injected by an outflow can be shared with a larger quantity of gas. This should have the effect of making outflow feedback far more effective in the presence of magnetic fields. This effect is demonstrated clearly in the simulations of \citet{Offner18a} in the context of exploring the effects of line-driven winds (as opposed to hydromagnetic winds) from intermediate mass stars on molecular clouds. They find that hydromagnetic waves that are launched from the working surfaces where winds impact molecular cloud material efficiently transfer energy and momentum over large distances, leading to significant turbulent motions far from the impact site.

Simulations bear out this conclusion. On the scales of individual cores with masses $\sim M_\odot$, \citet{Offner14b} and \citet{Offner17a} find that, for fixed outflow properties and initial conditions, a decrease in the mass to magnetic critical mass ratio from $\mu_\Phi = \infty$ to $\mu_\Phi = 1.5$ (corresponding to an increase from zero magnetic field to near-critical) is associated with a reduction in the fraction of mass accreted onto the final star from $\approx 50\%$ to $\approx 15\%$. Note, however, that this conclusion depends on the outflow properties being independent of the large-scale field, as is the case in \citet{Offner17a}'s simulations because the outflow launching region is not resolved, and thus the outflows are inserted by hand. In simulations with self-consistently launched outflows, \citet{Machida13a} find the opposite dependence, because stronger fields produce more magnetic braking, which in turn makes the outflows weaker. However, it is unclear how realistic this conclusion is, since \citeauthor{Machida13a}'s simulations use laminar initial conditions with well-ordered fields, and simulations with turbulent initial conditions and fields find that these greatly reduce the effectiveness of magnetic braking \citep{Santos-Lima12a, Seifried12a, Seifried13a}.

A similar dependence on magnetic fields is apparent in simulations of the formation of star clusters from gas clumps with masses of $\sim 100 - 1000$ $M_\odot$. For low-mass clusters, \citet{Hansen12a} found that outflows reduced the overall rate of star formation by a factor of $\sim 2$ in simulations that did not include magnetic fields, while for much more massive and dense clusters, \citet{Krumholz12b} found an even smaller reduction in $\eff$, by a factor of $\approx 1.2$. \citet{Murray18a} obtain a similarly-small effect. By contrast, simulations that include both outflows and magnetic fields find much stronger effects. \citet{Nakamura07a} and \citet{Wang10a} find that the combination of outflows plus magnetic fields yields a reduction in $\eff$ from $\approx 1$ to $\approx 0.1$ in clouds that are slightly magnetically supercritical. Moreover, the combination is sufficient to prevent the cloud from going into overall collapse, because outflow momentum coupled to the magnetic fields maintains the turbulent velocity dispersion, keeping the clouds near virial balance. \citet{Federrath2015} find that magnetic fields plus outflows together produce $\eff \approx 0.04$, which, given the systematic uncertainties discussed in \autoref{ssec:sfr_obs}, is within the range of the observations.

\begin{figure*}
\begin{center}
\includegraphics[width=8cm]{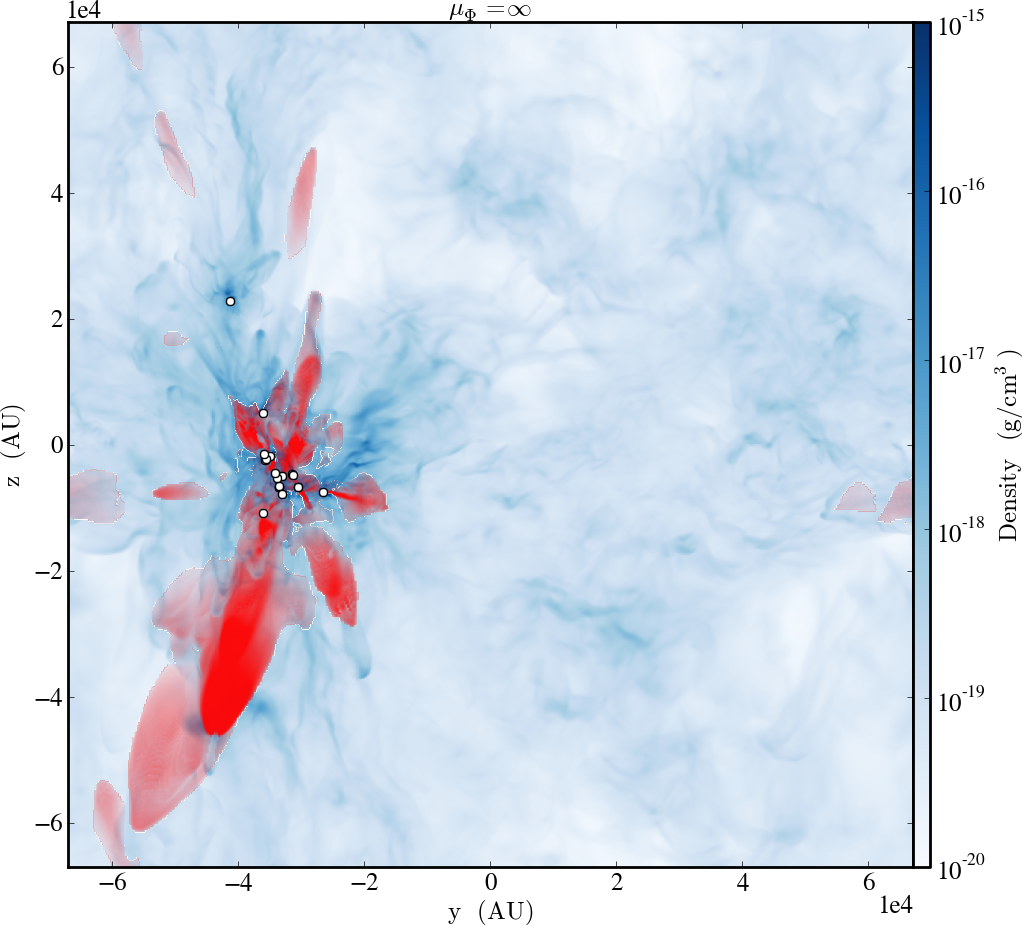}\hspace{0.2in}\includegraphics[width=8cm]{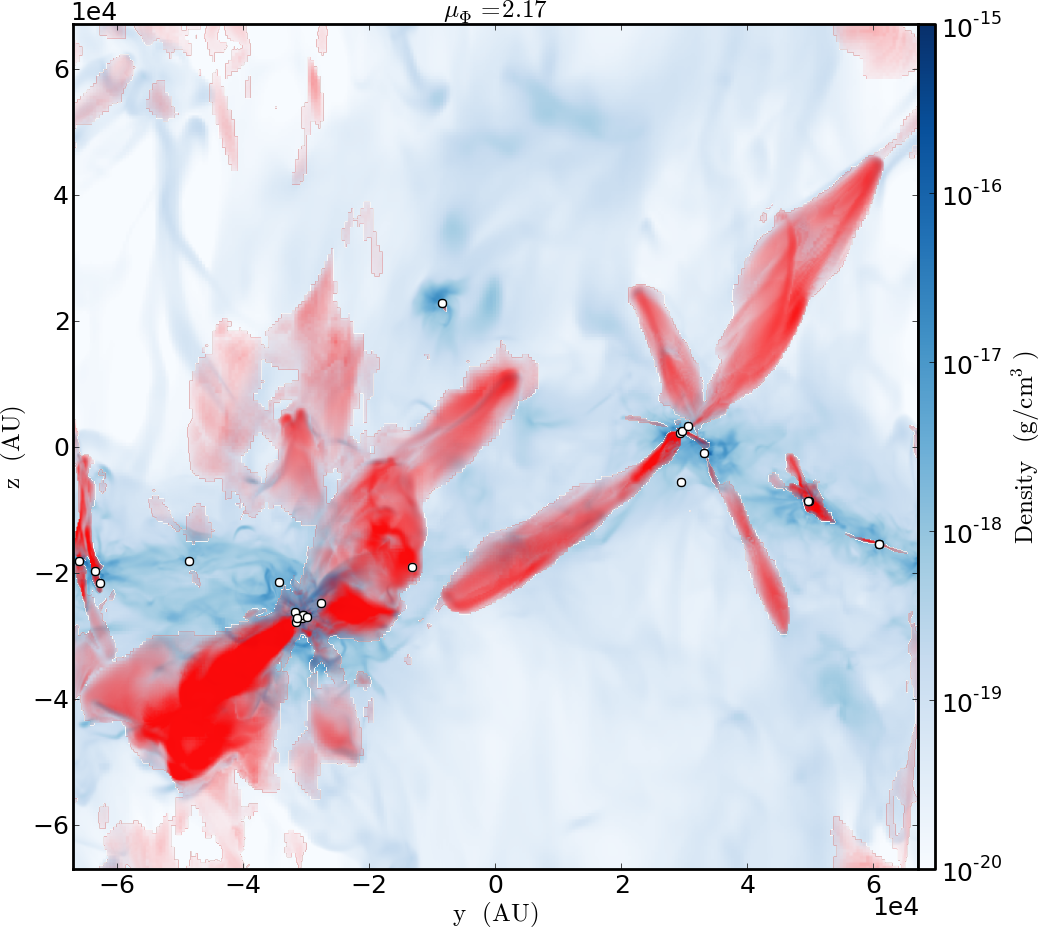}
\end{center}
\caption{
\label{fig:cunningham18}
Results from simulations of star cluster formation including outflows and radiative heating by \citet{Cunningham18a}, comparing a case without magnetic fields (\textbf{left}, mass to magnetic critical mass ratio $\mu_\Phi = \infty$) to an otherwise-identical simulation with a strong magnetic field (\textbf{right}, mass to magnetic critical mass ratio $\mu_\Phi = 2.17$). Blue color shows volume-weighted mean density projected along each line of sight. Red color indicates the presence of gas moving with velocity greater than twice the RMS speed in the simulation domain, with the opacity becoming complete at fives times the RMS speed; thus the red color mostly traces outflows or gas entrained by them. White circles indicate the positions of protostars. Note how the simulation without magnetic fields has most of the mass collapsed into a single dense clump, with outflows poking small holes but not ejecting much mass. The strongly magnetized run has a more distributed morphology, and outflows are more effective at preventing the build-up of dense structures.
}
\end{figure*}

In simulations including both radiative heating from young stars and outflows, \citet{Myers14a} find that the combination of these two effects in the absence of magnetic fields yields $\eff = 0.17$, while adding magnetic fields at a level corresponding to $\mu_\Phi = 2$ reduces this to $0.07$. \citet{Cunningham18a} obtain a similarly-large difference between runs with and without magnetic fields, which we illustrate in \autoref{fig:cunningham18}. Most recently, \citet{Li18a} have obtained $\eff \approx 0.03 - 0.07$ (depending on exactly how they measure it) in a simulation that self-consistently follow the formation and evolution of a cloud with radiative and outflow feedback.

In summary, magnetic fields appear to have a multiplicative effect on outflow feedback, producing a significantly greater reduction in $\eff$ than do either magnetic fields without outflows, or outflows without magnetic fields. Modern simulations that include both effects are now able to reproduce values towards the high end of the observed $\eff$ distribution. The remaining discrepancy may be due to other physical effects still missing in the simulations, or due to systematic errors at the factor of $\approx 3$ level affecting the observed $\eff$. There are systematic uncertainties on the values of $\eff$ from simulations as well, though these are likely somewhat smaller. For example, when measuring $\eff$ from a simulation, one must choose a Lagrangian region (e.g., all the mass above some density $\rho_{\rm min}$) or an Eulerian region (e.g., all the mass inside a simulation box) over which it is to be measured, and differences in how this region is chosen can lead to variations in the inferred $\eff$ value at the factor of $\approx 2$ level. Similarly, multiple simulations carried out with the same physical setup by different random realizations of turbulence show $\approx 50\%$ variations in $\eff$ \citep{Federrath12a}, though this issue has not been explored extensively in simulations including feedback due to their high computational cost.

\subsubsection{Photoionization}

While all forming stars likely produce outflows, only the most massive produce substantial ionizing luminosities. When such stars are present, however, they are probably the dominant sources of feedback at the scales of molecular clouds. Ionizing radiation heats the gas it encounters to temperatures $\approx 10^4$ K, such that the sound speed is $\approx 10$ km s$^{-1}$, well above the escape speed in most molecular clouds. Consequently, the ionized gas rapidly escapes from the cloud, directly removing mass and exerting back-forces on the remaining neutral material that can potentially drive turbulence or eject even more mass. The development of an H~\textsc{ii} region is the observable manifestation of this phenomenon, and both analytic models \citep[e.g.,][]{Matzner02a, Krumholz06a} and numerical simulations \citep[e.g.,][]{Grudic18a, Kim18a} suggest that H~\textsc{ii} region formation is important for regulating star formation rates in molecular clouds.

What role do magnetic fields play in these processes? \citet{Krumholz07a} provide a basic analytic outline, which they show roughly predicts the behavior of simulations. The ionized gas sound speed $c_i$ is much larger than the Alfv\'en speed $v_A$ in typical Galactic molecular clouds, so as an H~\textsc{ii} region begins expanding, the pressure of newly ionized gas is much larger than the magnetic pressure, and magnetic fields have little effect on the dynamics. As the ionized gas expands, however, its density drops, while the forces this gas exerts on neighboring neutral material cause it to compress, raising the magnetic field strength. Thus as H~\textsc{ii} regions evolve, the ionized gas pressure falls and the magnetic pressure and tension in the neighboring neutral material rise, until the forces become comparable. This occurs once the H~\textsc{ii} region reaches a characteristic size
\begin{eqnarray}
r_m & \equiv & \left(\frac{c_i}{v_A}\right)^{4/3} \left(\frac{3 Q}{4\pi \alpha_B f_e n_{\rm H,0}^2}\right)^{1/3}
\nonumber \\
& \approx & 1.6\, Q_{49}^{1/3} B_2^{-4/3} T_4^{0.94}\mbox{ pc}
\end{eqnarray}
where $c_i \approx 10$ km s$^{-1}$ is the ionized gas sound speed, $v_A$ and $n_{\rm H,0}$ are the Alfv\'en speed and number density of H nuclei in the undisturbed neutral medium into which the H~\textsc{ii} region is expanding, $Q$ is the ionizing luminosity measured in photons per unit time, $\alpha_B$ is the case B recombination coefficient, and $f_e$ is the mean number of free electrons per hydrogen atom in the ionized region. In the numerical evaluation we have adopted $f_e = 1.1$ (i.e., assumed He is singly-ionized), and defined $Q_{49} = Q/10^{49}$ s$^{-1}$, $B_2 = B/100$ $\mu$G, $T_4 = T/10^4$ K, with $T$ the temperature in the H~\textsc{ii} region; we evaluate $\alpha_B$ using the powerlaw approximation given by \citet{Draine11a}. We have chosen the numerical scalings so that all parameters are typically of order unity for an early O star and the magnetic field strengths typically observed towards regions of massive star formation \citep{Crutcher12a}.

Since the magnetic characteristic radius $r_m$ is smaller than the size of typical molecular clouds, magnetic forces will generally become non-negligible at some point during the evolution of a typical H~\textsc{ii} region. There is significant evidence for this from studies of H~\textsc{ii} region morphology. Simulations predict that significant magnetic forces cause H~\textsc{ii} regions to become elongated along the direction of the large-scale field, while the field is distorted into a ring-like morphology tracing the dense shell that forms the H~\textsc{ii} region's boundary \citep{Krumholz07a, Arthur11a, Wise11a, Mackey11a}. These features are in fact observed \citep{Pellegrini07a, Tang09a}. For example, \citet{Pavel12a} combine radio recombination line surveys for H~\textsc{ii} regions with near-IR polarimetry and find that young H~\textsc{ii} regions have their long axes preferentially aligned with the mean magnetic field of the galactic disk around them. \citet{Chen17a} measure the orientation of the magnetic field in the molecular gas ring N4, which traces the edges of an H~\textsc{ii} region, using near-IR polarimetry of background stars. They find that, exactly as the simulations predict, the magnetic field orientation on the plane of the sky is preferentially tangential to the ring, with 16/21 of the field orientation vectors lying within $30^\circ$ of this direction, and 10/21 lying within $10^\circ$.

\begin{figure*}
\begin{center}
\includegraphics[height=7cm]{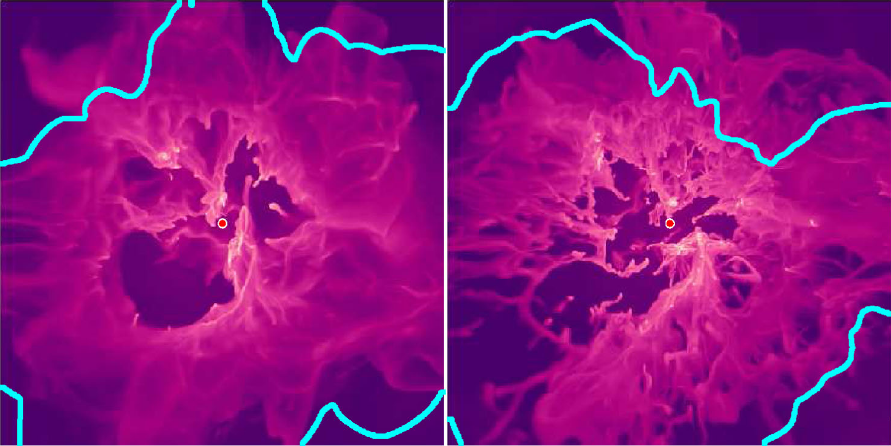}\hspace{0.1in}\includegraphics[height=7cm]{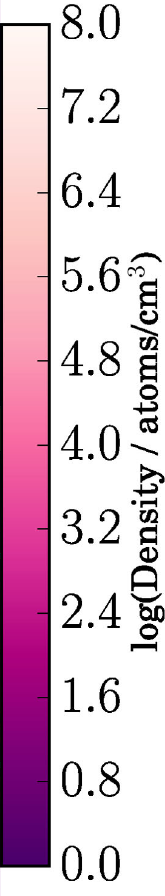}
\end{center}
\caption{
\label{fig:geen15}
Snapshots of two simulations by \citet{Geen15b}, one with a magnetic field (\textbf{left}) and one without (\textbf{right}). The region shown is a cube 27 pc on a side, with a single ionizing source with a luminosity $Q=10^{48}$ ionizing photons s$^{-1}$ located at the origin, indicated by the red circle. The slices show the state 2 Myr after the ionizing source turns on. Color shows the maximum density projected along each line of sight, as indicated by the color bar. The cyan contour marks the region where, somewhere along the line of sight, the ionization fraction exceeds 10\%. Notice how the presence of the magnetic field has prevented the H~\textsc{ii} region from blowing out.
}
\end{figure*}

It is less clear, however, whether magnetic effects are quantitatively important when it comes to determining the star formation rate. \citet{Gendelev12a} find that the compressed magnetic field associated with a magnetized H~\textsc{ii} region stores a significant energy reservoir, which at least has the potential to drive motions and convert a greater fraction of the injected energy to turbulence than would be the case for a non-magnetized region. While the latter effect has yet to be demonstrated in simulations of H~\textsc{ii} regions, the analogous process has been demonstrated for wind feedback by \citet{Offner18a}. \citet{Geen15b, Geen17a} find that magnetic fields help confine H~\textsc{ii} regions and prevent gas and ionizing photons from escaping; we reproduce two snapshots from their simulations in \autoref{fig:geen15}. However, this effect changes the total H~\textsc{ii} region energy and momentum budget relatively little, suggesting that the impact on star formation (which is not included in their simulations) might also be relatively small. To date there have been far fewer systematic studies of the interaction of photoionization feedback with magnetic fields than for outflow feedback, and thus the range of possible effects is much less certain.

\subsubsection{Supernovae, Winds, and Interface Mixing}

Supernovae (SNe) represent another form of feedback with which it is possible for magnetic fields to interact. While photoionization is the dominant form of feedback on the scales of molecular clouds, SNe are more important at galactic scales, and in the past few years a number of authors have argued that either the large-scale rate of star formation in galaxies, the velocity dispersion of the ISM on large scales, or both, are ultimately dictated by the amount of radial momentum injected into the ISM when a SN explosion occurs \citep[e.g.,][]{Dib06a, Joung06a, Hopkins11a, Ostriker11a, Shetty12a, Faucher-Giguere13a, Krumholz18a}. For a single SN, many authors have found that this radial momentum budget is $\approx 3\times 10^5$ $M_\odot$ km s$^{-1}$ per SN \citep[e.g.,][]{Iffrig15a, Kim15a, Martizzi15a, Walch15a}, and theoretical models for the ISM often adopt this value.

Since the Alfv\'en speed in galactic disks is far less than the expansion speeds of SN remnants (SNRs), at least until very late in their evolution, magnetic forces are generally unimportant for SNRs on large scales; this makes them fundamentally different than H~\textsc{ii} regions, where magnetic forces become important early on. However, magnetic fields may nevertheless play an important role on small scales. The dynamics of a SNR, particularly one driven by multiple SNe occurring over time, are ultimately controlled by the rate of radiative energy loss from the SN-heated gas that acts as a piston to drive the expansion of the surrounding cold ISM; the longer it takes the hot gas to radiate away its energy, the more energy is available to accelerate the cold ISM.\footnote{The arguments about SNRs that we make here apply equally well to bubbles of hot gas produced by the radiatively-driven winds of massive stars. We focus on SNRs because they are likely more important for regulating star formation rates, but the underlying physical issues are much the same for wind bubbles.} This energy loss, in turn, has the potential to be completely dominated by the interface layer between the hot and cold fluids, and thus the rate of energy loss depends critically upon the rate of mixing across the contact discontinuity separating hot and cold gas \citep{McKee84a, Tenorio-Tagle90a, Tenorio-Tagle91a, Strickland98a}. Differing assumptions about the rate of mixing lead to order of magnitude or larger variations in the predicted X-ray luminosities of hot bubbles \citep[e.g.,][]{Dunne03a, Rosen14a}, with corresponding variations in the amount of momentum that an expanding hot bubble can deliver before radiative cooling saps its energy \citep{Keller14a, Keller15a, Fierlinger16a, Gentry17a}. 

This is not a small effect: for example, \citet{Gentry17a} survey a large parameter space of supernova number, metallicity, and ISM density using 1D simulations, and find that, if there is negligible mixing across the interface, a SNR driven by a cluster of 10 SNe will on average inject $\approx 10$ times as much radial momentum \textit{per SN} (i.e., about $3\times 10^6$ $M_\odot$ km s$^{-1}$ per SN instead of $3\times 10^5$) as a SNR driven by a single star. \citet{Sharma14a} and \citet{Yadav17a} find similarly-large enhancements from clustering in their 3D simulations of a smaller parameter space. Averaging of the star cluster mass function, \citet{Gentry17a} find a net increase in momentum yield per SN of a factor of $\approx 4$ compared to the commonly-adopted value. On the other hand, if there is efficient mixing, then clustering of SNe does not substantially change the momentum budget. Depending on the large-scale ISM model adopted, this factor of $\approx 4$ variation in the SN momentum budget implies either a factor of $\approx 4$ variation in the star formation rate, the ISM velocity dispersion, or some combination of the two.  Consequently, any mechanism that alters the rate of mixing across contact discontinuities between hot and cold gas has the potential to alter the effects of SN feedback on the structure and star formation rate of the ISM at this level.

Magnetic fields potentially play an important role in this problem because they suppress mixing across contact discontinuities, and thus tend to push toward higher momentum yields from SNRs. This suppression takes two forms: first, magnetic fields prevent electrons from free-streaming across field lines, which tend to be parallel to the contact discontinuity as a result of sweeping-up of pre-existing fields by the expanding hot bubble; this greatly reduces the rate of thermal conduction \citep[e.g.,][]{Vikhlinin01a, Markevitch07a}. Second, by providing a surface tension-like force, magnetic fields parallel to an interface strongly suppress physical mixing between two fluids by suppressing instabilities such as Rayleigh-Taylor and Kelvin-Helmholtz that would otherwise mix fluids \citep[e.g.,][]{Stone07a, Stone07b, McCourt15a, Banda-Barragan16a, Banda-Barragan18a}. \citet{Offner15a} find that this effect can be partially offset by magnetic kink instabilities, but the net amount of mixing across the interface is still reduced by the presence of a field. In direct simulations of SNR expansion, \citet{Gentry19a} find that, at fixed resolution, simulations including magnetic fields lead to SNRs having noticeably larger terminal radial momenta than purely hydrodynamic simulations.

However, the magnitude of this effect remains very poorly-understood due to the extremely challenging numerics of the problem. To obtain a result for the terminal momentum of a SNR one must of course simulate its full expansion, which will easily reach size scales of a few hundred pc if there are multiple SNe. However, one must simultaneously resolve the edge of the SNR well enough that numerical mixing does not dominate the transport rate across the contact discontinuity. The characteristic thickness of the interface, set by balancing the rate of conductive heat flux from hot to cold against the rate of radiative loss, is the Field length \citep{Field65a, Koyama04a},
\begin{equation}
\lambda_{\rm F} = \sqrt{\frac{\kappa_c T}{n_{\rm H}^2\Lambda}},
\end{equation}
where $\kappa_c$ is the conduction coefficient, $T$ is the temperature, $n$ is the number density of H nuclei, and $\Lambda$ is the cooling function (i.e., the energy radiated per unit volume per unit time is $n_{\rm H}^2\Lambda$). The conductivity, assuming the unsaturated limit and a gas of fully ionized H and He in the usual interstellar ratio, is \citep{Cowie77a}
\begin{equation}
\kappa_c \approx \frac{1.84\times 10^{10} T_6^{5/2}}{29.9+\ln (T_6 n_0^{-1/2})}\mbox{ erg s}^{-1}\mbox{ K}^{-1}\mbox{ cm}^{-1},
\end{equation}
while in the temperature range $\sim 10^4 - 10^6$ K that characterizes the interface, the cooling rate for Solar metallicity gas is \citep{Draine11a}
\begin{equation}
\Lambda \approx 1.3 \times 10^{-22} T_6^{-0.7}\mbox{ erg cm}^3\mbox{ s}^{-1},
\end{equation}
where $T_6 = T/10^6$ K and $n_0 = n_{\rm H}/1$ cm$^{-3}$. For $n_{\rm H} = 0.1$ cm$^{-3}$ and $T = 10^5$ K, typical interface values, we have $\lambda_{\rm F} \approx 0.05$ pc, implying that effective resolutions of $>1000^3$ would be required to capture the interface and the SNR as a whole in the same simulation.

Not surprisingly, numerical simulations have struggled to reach this goal. Without magnetic fields, \citet{Fierlinger16a} are able to obtain convergence in their 1D Eulerian simulations only if they impose a subgrid diffusion model that corresponds to assuming efficient turbulent mixing across the contact discontinuity. \citet{Gentry17a} do obtain convergence in their 1D simulations of SNR evolution with multiple SNe without such a model, but only using a pseudo-Lagrangian method to minimize numerical mixing across the hot-cold interface, and upon reaching a resolution $\Delta x \approx 0.03$ pc; they are unable to obtain convergence with Eulerian methods. In 3D hydrodynamic simulations, \citet{Yadav17a} and \citet{Gentry19a} find that SNR energies and momenta are still not converged at resolutions of a few tenths of a pc, the highest they could simulate. In contrast, \citet{Kim17a} do report convergence in their 3D simulations at a factor of several lower resolution, 1.5 pc, which they attribute to the fact that they simulate a non-uniform background into which the SNR expands, and that this makes convergence easier to obtain. \citet{Gentry19a}, on the other hand, suggest that the convergence might instead be an artifact of mixing being dominated by the advection of the contact discontinuity across the grid, which might not converge as the resolution increases, since the front would mix less per cell but would have to cross a larger number of cells per unit time. In summary, we are still some distance from determining the true momentum of SNRs even in the hydrodynamic case. It seems unlikely we will be able to measure the difference between this case and the magnetized one until we make progress on issues of convergence.

\subsubsection{Cosmic Ray Feedback}

The final form of feedback that we discuss is cosmic rays (CRs). A full review of CR physics is well beyond the scope of this review, and we refer readers to \cite{Zweibel13a} for a comprehensive treatment; here we only summarize the most important features. CRs are a population of non-thermal particles created when charged particles bounce back and forth across magnetized shocks; each passage through the shock increases the particle energy, allowing the shock to act like a particle accelerator. Magnetic fields are required to create CRs, but they are also critical for providing a mechanism by which CRs can couple to gas dynamics: CRs scatter off Alfv\'en waves or other inhomogeneities in magnetic fields, transferring momentum in the process. Thus CR feedback is fundamentally a magnetic process.

One critical question for CR feedback is the size scale on which it is effective. While any magnetized shock in a sufficiently-ionized plasma can accelerate CRs, the bulk of the CR energy budget on galactic scales comes from SN shocks, which convert $\sim 10\%$ of their initial kinetic energy into CRs. This population is injected on the scales of SN remnants, which are much larger than individual molecular clouds, and the population further spreads out in height as it diffuses through the galactic magnetic field. Thus while the pressure provided by CRs at the midplane of the Milky Way or similar galaxies is comparable to the magnetic or turbulent ram pressures, the scale height of the CRs is much larger than that of the star-forming molecular gas \citep{Boulares90a}. For this reason, most recent work on CR feedback has focused on their possible role as drivers of galactic winds \citep[e.g.,][among many others]{Uhlig12a, Girichidis18a} or sources of heating in galaxy winds or halos \citep[e.g.,][]{Wiener13a, Ruszkowski17a}, in which role they would affect star formation only indirectly, but modulating the fuel supply for it. It is unclear if CRs can affect the SFR for gas already in a galaxy. \citet{Socrates08a} suggest that CR feedback prevents galactic SFRs per unit area from exceeding some maximum value. While observations do suggest that there is in fact an upper limit to galaxy areal SFRs, CRs are far from the only possible explanation for it \citep[e.g.,][]{Crocker18a}, and the \citeauthor{Socrates08a} calculation is not precise enough to allow quantitative comparison to the observations.

On the smaller scales of individual molecular clouds, for CR feedback to be dynamically significant there must be some mechanism for producing a CR pressure gradient.\footnote{An important distinction to draw here is between CRs providing a dynamically important pressure, and being important in other ways. CRs are certainly critical to the ionization state, temperature, and chemistry of molecular gas, even if they are not dynamically important.} One potential mechanism for producing a gradient is absorption of low-energy, non-relativistic CRs by molecular gas. Clouds with column densities $\gtrsim 10^{23}$ cm$^{-2}$ dissipate CRs with the streaming instability, ultimately converting much of the CR energy to turbulent motions \citep{Schlickeiser16a}; to date there has been no exploration of whether this could be a significant source of turbulence in molecular clouds, that \textit{a priori} it seems unlikely on energetic grounds, since the energy density of CRs at a galactic midplane is comparable to the mean kinetic energy density, while the kinetic energy density associated with turbulence in a molecular cloud is $\sim 2$ orders of magnitude larger. One can make a similar point about another possible source of inhomogeneity, CRs generated by protostellar jets \citep{Padovani15a}: while these may be important sources of ionization, even if one assumes efficient CR acceleration such that $\approx 10\%$ of the energy in jets is ultimately transferred to CRs, this is not enough to be dynamically significant compared to the binding energy of an entire molecular cloud. CRs accelerated in shocks from the winds of massive stars are a more promising potential origin for a locally-inhomogeneous CR population, since the associated energy budget is considerably larger. CRs created in such shocks are likely sub-dominant but non-negligible on galactic scales \citep{Seo18a}, but there is significant observational evidence that the CR population these generate is localized around massive star clusters (see the review by \citealt{Bykov14a}), and thus could potentially provide a dynamically-significant outward pressure that would lower SFRs. This possibility has yet to receive significant theoretical or observational attention.

\section{The Role of Magnetic Fields for the Initial Mass Function}

\subsection{Basics of the IMF and Observational Evidence}

Extensive general reviews of the IMF -- in particular the observational challenges involved in measuring the IMF -- are provided by \citet{OffnerEtAl2014} and \citet{Hopkins2018}. Here we concentrate on the effects of magnetic fields on the IMF. The IMF is the distribution of stellar masses at birth. We know from observational surveys that most stars have masses of about half the mass of our Sun ($\msol$). Stars with smaller masses are rarer. Stars more massive than the Sun also become rarer with increasing mass. The high-mass tail of the IMF is indeed a steeply decreasing power-law function with the number of stars $N(M)\propto M^{-1.35}$ \citep{Salpeter1955,MillerScalo1979,Kroupa2001,deMarchiParesce2001,Chabrier2003,Chabrier2005,ParravanoMcKeeHollenbach2011,ParravanoHollenbachMcKee2018,Da-Rio12a,Weisz15a}.

\begin{figure}
\includegraphics[width=\columnwidth, trim=0.25cm 0.0cm 0.25cm 0.0cm, clip=true]{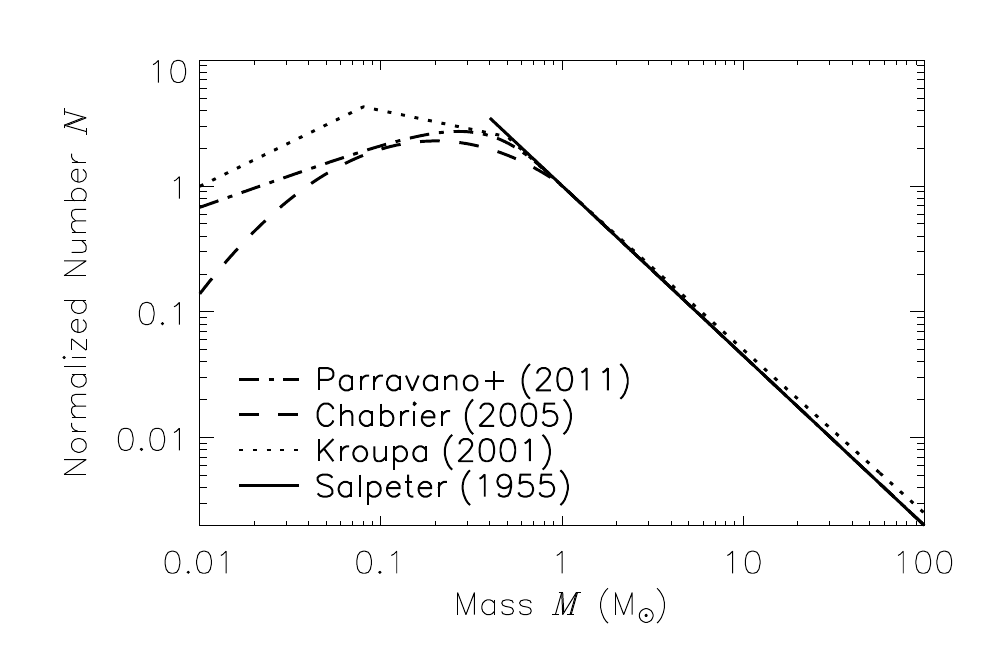}
\includegraphics[width=\columnwidth, trim=0.25cm 0.0cm 0.25cm 0.0cm, clip=true]{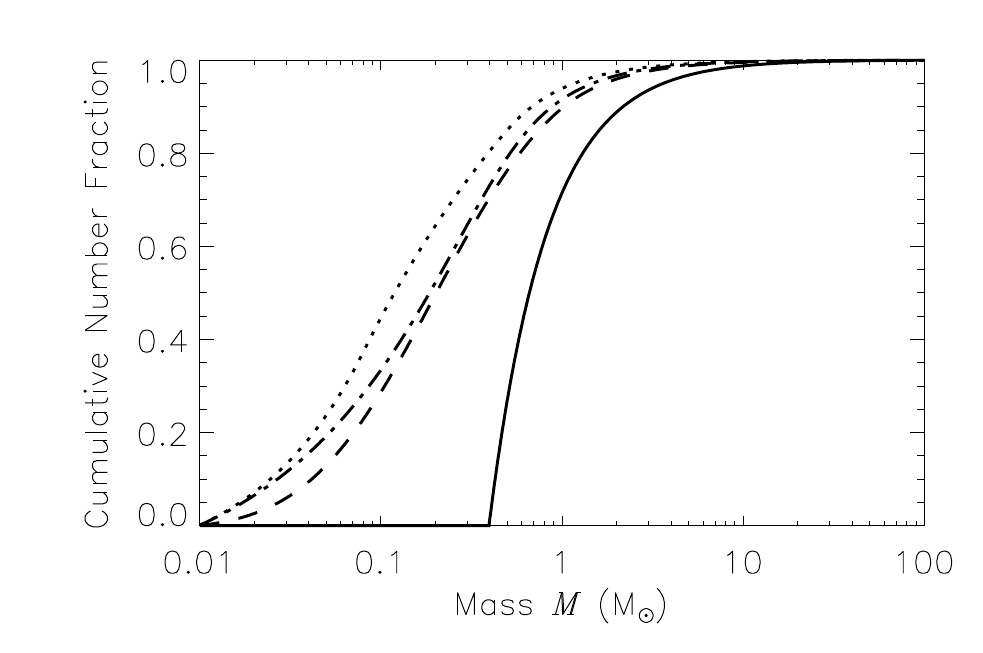}
\caption{Analytic fits to the observed IMF (\textbf{top}) and the cumulative mass function of stars (\textbf{bottom}). Different lines show different parameterizations by \citet{Salpeter1955} (solid), \citet{Kroupa2001} (dotted), \citet{Chabrier2005} (dashed), and \citet{ParravanoMcKeeHollenbach2011} (dash-dotted). In this representation of the IMF, the number of stars is normalized such that $N=1$ for $M=1\,\msol$. While the high-mass tail ($M\gtrsim1\,\msol$) seems fairly universal, the low-mass end ($M\lesssim1\,\msol$) is much less well constrained, with substantial variations in the number of low-mass stars and in the characteristic mass of the IMF.}
\label{fig:imf}
\end{figure}

\autoref{fig:imf} shows a compilation of various analytic fits to the observed IMF. There is clearly substantial disagreement on the low-mass end ($M\lesssim1\,\msol$) with the turnover mass (or characteristic mass) varying between $\sim0.1\,\msol$ and $\sim0.4\,\msol$ depending on the parameterization. This disagreement is a result of the challenges in observing low-mass stars, taking into account multiplicity, and converting from a luminosity function to a mass function \citep{OffnerEtAl2014,Hopkins2018}. For the high-mass tail ($M\gtrsim1\,\msol$), however, there seems to be generally good agreement.

Efforts to search for systematic variation in the IMF have yielded mixed and often contradictory results. In the Milky Way, \citet{Weidner13a}, \citet{Dib2014}, and \citet{DibSchmejaHony2017} argue that there is statistically-significant evidence for variation in both the low-mass and high-mass parts of the IMF from one star cluster to another. However, as pointed out by \citet{Krumholz14a}, the quoted uncertainties in these studies frequently ignore the two largest systematic uncertainties: errors in stellar masses derived from pre-main sequence tracks, and in errors in the masses and other properties of star clusters that are simply drawn from the literature, and rather than derived using homogeneous and uniform cluster definitions or analysis methods. Searches for IMF variation using homogeneous samples in external galaxies have for the most part found no statistically-significant variation at least at the high-mass end of the IMF that is accessible beyond the Milky Way \citep[e.g.,][]{Andrews13a, Andrews14a, WeiszEtAl2015}. The main exceptions are in the most massive star clusters, where \citet[in 30 Doradus]{SchneiderEtAl2018Sci} and \citet[in the Arches Cluster]{Hosek19a} have reported statistically-significant excesses of massive stars compared to the average IMF of the Galactic field. There is also more indirect evidence for bottom-heavy IMFs in massive elliptical galaxies (see the review by \citealt{Hopkins2018}). Given the highly uncertain status of observational searches for IMF variation, and the fact that at this point there is no reason to think any variations that might exist are linked to magnetic fields, we will not discuss this topic further.

Understanding the power-law tail in the IMF and the turnover at around \mbox{$0.1$--$0.4\,\msol$} are two of the most challenging open problems in astrophysics. The IMF has far-reaching consequences and applications, including the calibration of extra-galactic star formation relations used to understand galaxy formation and evolution \citep{GreenEtAl2010}. The IMF is needed to interpret the colors, brightness and star formation activity of all galaxies in our Universe and it is the central ingredient for understanding galaxy formation and evolution, because the feedback from young stars is what powers the life cycle of galaxies.

Many physical processes may play a role in setting the characteristic mass and shape of the IMF, including gravity, turbulence, magnetic fields, and feedback, as proposed in theoretical models and seen in numerical simulations. However, we are not aware of any direct observational test of these theoretical predictions, especially when it comes to the role of magnetic fields for the IMF. Given the lack of observational constraints, we thus need to resort to theoretical models and numerical simulations to advance our understanding of the physical mechanisms that control the IMF.

\subsection{Theoretical Models of the IMF}

\subsubsection{Magnetic Jeans Mass} \label{sec:mj}
Analytic work on the effects of magnetic fields for the IMF are scarce. The earliest and simplest approaches to incorporating magnetic fields into theories of the IMF simply assumed that fields would convert the geometry from spherical to filamentary, and then proceeded to calculate a Jeans length or mass in the resulting geometry, neglecting any further magnetic effects \citep[e.g.,][]{Inutsuka01a, Larson2005}. A slightly more sophisticated approach is to invoke a magnetic version of the Jeans length,
\begin{eqnarray}
\lambda_\mathrm{J,mag} & = & \left[\frac{\pi\cs^2\left(1+\beta^{-1}\right)}{G\rho}\right]^{1/2}
\nonumber \\
& =& \lambda_\mathrm{J} \left(1+\beta^{-1}\right)^{1/2}\,,
\label{eq:mjb}
\end{eqnarray} 
which leads to the magnetic Jeans mass
\begin{equation} \label{eq:mj}
M_\mathrm{J,mag} = \rho\frac{4\pi}{3}\left(\frac{\lambda_\mathrm{J,mag}}{2}\right)^3 = M_\mathrm{J} \left(1+\beta^{-1}\right)^{3/2}\,,
\end{equation}
where $\lj$ and $\mj$ are the standard (purely thermal) Jeans length and mass, respectively. All we have done here is to replace the thermal pressure with the sum of thermal and magnetic pressure, giving rise to the $(1+\beta^{-1})$ correction factors \citep{Federrath12a,Hopkins13a}, introducing the plasma $\beta$ in the relations. This simple concept shows that adding magnetic pressure raises the Jeans mass. If the Jeans mass plays a role in setting the characteristic mass of stars \citep{OffnerEtAl2014}, then \autoref{eq:mjb} would suggest that adding magnetic pressure leads to more massive stars (or less fragmentation). For example, taking a typical value of $\beta=0.3$ for molecular clouds leads to an increase compared to the purely thermal Jeans mass by a factor of $\sim9$. We caution that this calculation is solely based on adding the magnetic pressure contribution to the Jeans mass, but ignores any potential effects of magnetic tension. These limitations have been discussed in \citet{Molina12a}, \citet{Federrath12a} and \citet{FederrathBanerjee2015}.

\subsubsection{MHD Turbulence-Regulated IMF Theories} \label{sec:imftheory}
The structure and dynamics of molecular clouds and dense cores are largely determined by MHD turbulence \citep{ElmegreenScalo2004,MacLowKlessen2004,McKeeOstriker2007}, and this MHD turbulence may not only control the rate of star formation, but also the mass of young stars. In the relevant context of magnetic fields, \citet{PadoanNordlund2002} presented a theory of the IMF for which the density PDF and the turbulence power spectrum are the main ingredients, complemented by the MHD shock jump conditions. Assuming that the density contrast in an MHD shock is proportional to the Alfv\'en Mach number, i.e., $\rho'/\rho\propto\macha$ and the post-shock thickness $\ell'/\ell\propto\macha^{-1}$, combined with the velocity dispersion -- size relation, $v \propto \ell^p$ with \mbox{$p\sim0.4$--$0.5$} from observations \citep{Larson1981,SolomonEtAl1987,OssenkopfMacLow2002,HeyerBrunt2004,RomanDuvalEtAl2011} and numerical simulations \citep{KritsukEtAl2007,SchmidtEtAl2009,FederrathDuvalKlessenSchmidtMacLow2010,Federrath2013}, they derive a model for the high-mass tail of the IMF,
\begin{equation}
N(M) \propto M^{-3/(3-2p)},
\end{equation}
which, for \mbox{$p=0.4$--$0.5$}, gives high-mass slopes of $-1.4$ to $-1.5$ for the IMF, very close to the observed \citet{Salpeter1955} slope. This slope is also consistent with the distribution of clump masses obtained in MHD turbulence simulations by \citet{PadoanEtAl2007}, though the simulations did not include gravity.

A significant problem with the theoretical model by \citet{PadoanNordlund2002} is that it needs a linear scaling of post-shock density and post-shock thickness with Mach number, as assumed above. However, MHD turbulence simulations with realistic values of the magnetic field show that the density contrast in shocks is not reduced by as much in the presence of magnetic fields as assumed in \citet{PadoanNordlund2002}. In fact, the more appropriate and effective scalings of post-shock density and thickness may be $\rho'/\rho\propto\mach^2$ and $\ell'/\ell\propto\mach^{-2}$, in which case the same derivation leads to
\begin{equation}
N(M) \propto M^{-3/(3-4p)},
\end{equation}
for the high-mass tail, significantly too steep, i.e., with slopes of $-2.1$ to $-3.0$, much steeper than the observed Salpeter slope.

\begin{figure*}
\centerline{\includegraphics[width=1.0\linewidth]{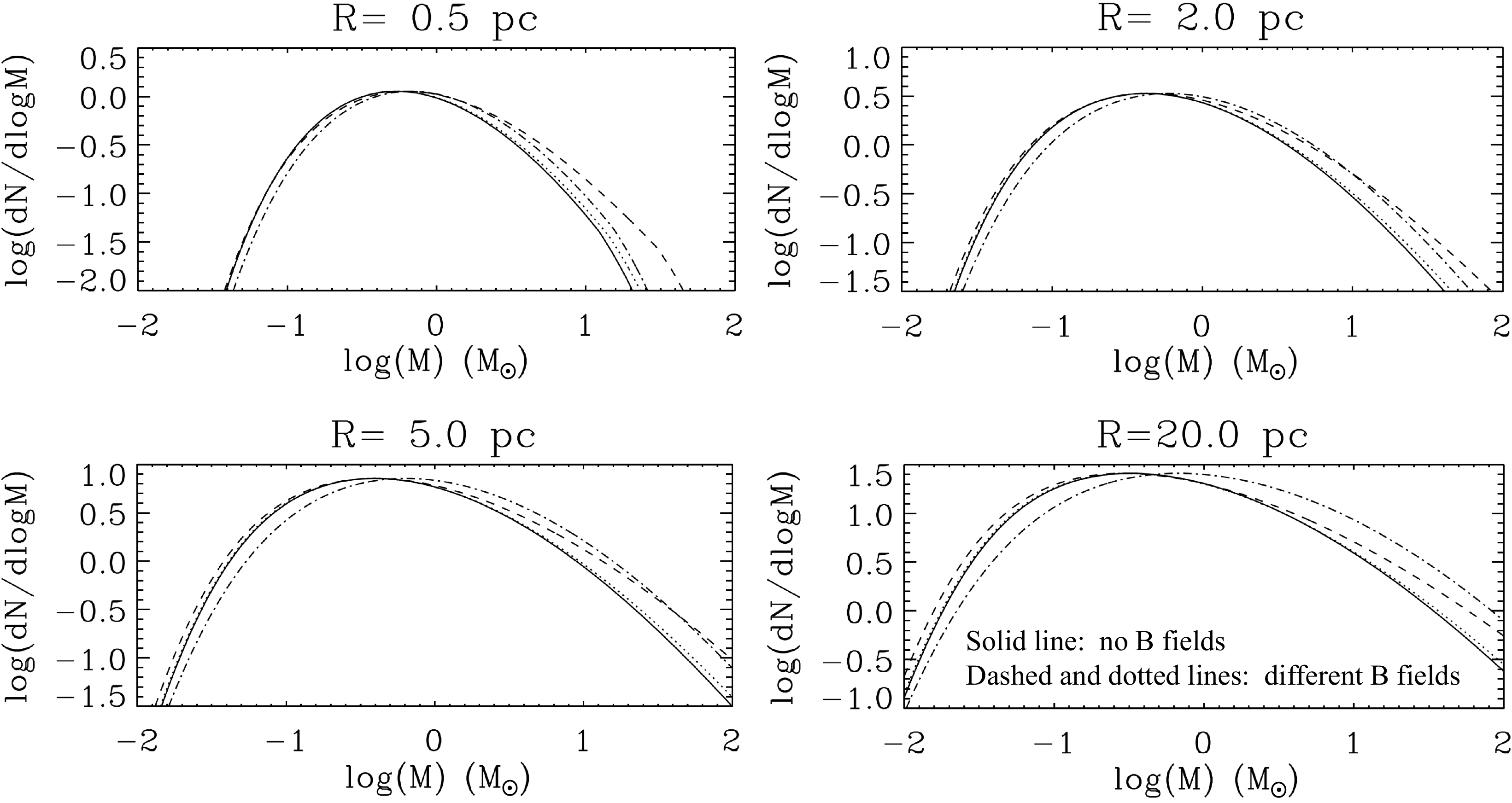}}
\caption{Analytic predictions of the core mass function with and without magnetic fields for different cloud radii, \mbox{$R=0.5$--$20\,\pc$}. The solid line is for the case without magnetic fields. The dotted line is for $B=10\,\mu\Gauss\,(n/(1000\,\cm^{-3}))^{0.1}$, the dashed line for $B=30\,\mu\Gauss\,(n/(1000\,\cm^{-3}))^{0.1}$, and the dash-dotted line for $B=10\,\mu\Gauss\,(n/(1000\,\cm^{-3}))^{0.3}$, where $n$ is the molecular hydrogen number density. Figure adopted from \citet{HennebelleChabrier2013}.
}
\label{fig:hc13}
\end{figure*}

\citet{HennebelleChabrier2008,HennebelleChabrier2009,HennebelleChabrier2013} present a similar class of turbulence-regulated models of the IMF, based on the \citet{PressSchechter1974} formalism, that yields IMF predictions in good general agreement with the observed IMF. \autoref{fig:hc13} shows the effect of adding magnetic fields in their model. Here we show predictions for the core mass function (CMF), i.e., a distribution that \citet{HennebelleChabrier2013} take to be shifted to three times higher masses compared to the IMF. We will comment further on the shift between the CMF and IMF in \autoref{sec:coretostareff}, which might be the result of magnetic-field driven outflow feedback.

We see in \autoref{fig:hc13} that the effect of the magnetic field (dashed and dotted lines for different magnetic field normalizations and scalings, bracketing the observed ranges) is relatively weak, when compared to the predictions without magnetic fields (solid lines). The magnetic field generally increases the characteristic mass of the IMF, consistent with the qualitative trend predicted simply by considering the magnetic Jeans mass (c.f.~\autoref{sec:mj}), but by much less than the factor of $\sim9$ based on \autoref{eq:mj}.

Looking in more detail at \autoref{fig:hc13}, we see that a stronger dependence of $B$ on the gas density (dash-dotted line) produce a stronger shift toward larger masses and stronger magnetic field normalizations (dashed line) yield a shallower slope in the high-mass tail. Both effects are the result of increased magnetic support, i.e., the addition of magnetic pressure to thermal pressure. These direct predictions by the \citet{HennebelleChabrier2013} theory of how the IMF would respond to different magnetic field strengths and field scalings with density have so far not been tested with numerical simulations.

In contrast, the role of magnetic fields in the analogous \citet{Hopkins13a} model is that they are degenerate with other parameters, i.e., any change in the IMF induced by magnetic fields could be reproduced by a change in Mach number, turbulence driving parameter or adiabatic index $\gamma$. Thus, in these models, magnetic fields do not have distinct effects that could be isolated from variations in other parameters.

\subsection{Numerical Simulations of the IMF}

Numerical simulations find that the overall effect of magnetic fields is to reduce the fragmentation of the gas. This is seen in both molecular cloud simulations \citep{PriceBate2008,Dib10a,Padoan11a,Federrath12a,Federrath2015} and protostellar disk simulations \citep{PriceBate2007,HennebelleTeyssier2008,BuerzleEtAl2011,PetersEtAl2011,HennebelleEtAl2011,SeifriedEtAl2011}. The physical reason for this is a combination of magnetic pressure and tension forces, the former with the effect of reducing compression, thereby increasing the effective Jeans mass (c.f.~\autoref{sec:mj}), and the latter acting to keep together coherent filaments, gas streams, and shocks by magnetic tension. These effects tend to produce less fragmented, more massive dense cores when magnetic fields are included. If this direct effect of magnetic fields on the gas were the only relevant effect, we would expect magnetic fields to increase the characteristic mass of stars compared to the purely hydrodynamical case.

However, the situation is slightly more complicated, because magnetic fields are the main reason for mechanical feedback in the form of jets and outflows launched from the accretion disk around young stars \citep{PudritzEtAl2007,FrankEtAl2014}. This jet/outflow feedback is also the reason why simple considerations based on magnetic Jeans mass (c.f.~\autoref{sec:mj}), and the more sophisticated models presented in \autoref{sec:imftheory}, may ultimately fail when it comes to the effect of magnetic fields. These models do not include feedback -- at least not its non-linear effect, which can ultimately only be properly accounted for and quantified in fully three-dimensional, MHD calculations. Jet/outflow feedback may be particularly important because it is the first to kick in (before radiation feedback, winds, and supernovae) and is not only important for high-mass stars, but applies to all young stars \citep{Krumholz14a}. Radiation feedback may also play an important role in determining the IMF, and we discuss the interplay between it and magnetic fields in \autoref{ssec:mag_rad}.

\subsubsection{Mechanical Feedback by Magnetically-Driven Jets and Outflows} \label{sec:coretostareff}

\paragraph{The Core-to-Star Efficiency}

\begin{figure*}
\centerline{\includegraphics[width=\columnwidth]{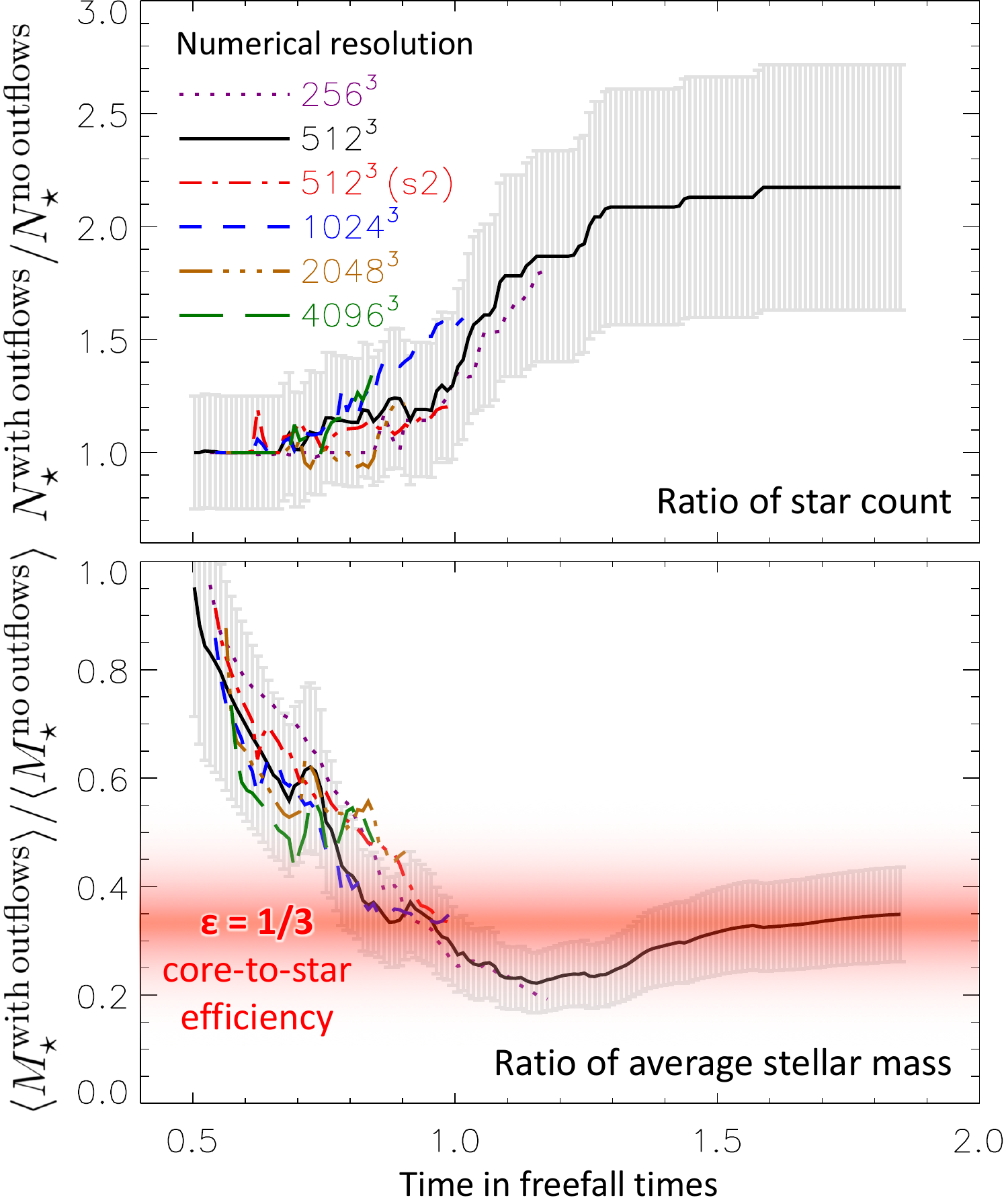}}
\caption{Time evolution of the ratio of the number of sink particles formed in simulations with and without magneto-centrifugal outflows, $N_{\star}^\mathrm{with\,outflows} / N_{\star}^\mathrm{no\,outflows}$ (top), and ratio of the average sink particle mass $\langle M_\star^\mathrm{with\,outflows} \rangle / \langle M_\star^\mathrm{no\,outflows} \rangle$ (bottom). Different lines show different numerical resolutions, demonstrating convergence. After a freefall time, outflow feedback has increased the number of sink particles formed by a factor of $\sim1.5$. The average sink particle mass is reduced by a factor of $\sim3$ with outflow feedback. Figure adopted from \citet{FederrathEtAl2014}.}
\label{fig:f14}
\end{figure*}

In the first part of this review we have seen that magnetically-driven outflows can reduce the SFR by factors of 2--3 and set the core-to-star efficiency to about 1/2. We therefore expect a significant impact also on the characteristic stellar mass and the IMF. Previous simulations have quantified this effect. For example, \citet{Hansen12a} found a reduction of the average stellar mass when outflow feedback was included in their simulations. Similarly, \citet{FederrathEtAl2014} observed additional fragmentation with outflow feedback. This is shown in \autoref{fig:f14}, where we plot the number of sink particles formed in simulations with outflow feedback divided by the number of sink particles formed in runs without magnetically-driven outflows. \autoref{fig:f14} shows that $N_{\star}^\mathrm{with\,outflows} / N_{\star}^\mathrm{no\,outflows} \sim 1.5$ after one freefall time. This is the result of outflow-induced fragmentation; the outflows perturb and tear filamentary accretion flows, breaking them up into multiple new accretion streams. Similar behavior has been observed in earlier simulations by \citet[][]{Wang10a}, \citet{LiEtAl2010} and \citep{Hansen12a}.

Magnetically-driven outflow feedback has two important effects on the stellar mass. First, it reduces the accretion rate and limits the final star mass by removing gas from the feeding core, leading to a core-to-star efficiency of $\sim0.5$. Second, it promotes fragmentation of the core, because the outflows tear up coherent accretion streams and perturb the core, such that more stars can form. This combined effect of magnetic outflows on the average stellar mass is shown in the bottom panel of \autoref{fig:f14}, which plots the ratio of the average stellar mass with and without outflow feedback, $\langle M_\star^\mathrm{with\,outflows} \rangle / \langle M_\star^\mathrm{no\,outflows} \rangle $. Comparing simulations with and without outflows, the mean stellar mass is the same at early times, immediately after the first collapsed objects form. However, stars grow more quickly in the simulations without outflows, so that after one free-fall time the mean stellar mass is a factor of $\sim 3$ smaller in simulations that include outflows. This factor of $\sim 3$ reduction in the final stellar mass is consistent with the results of other simulations \citep{LiEtAl2010,Hansen12a,Myers14a,Offner17a,Cunningham18a}. This suggests that magnetically-driven outflows may play a crucial role in controlling the observed shift of the core mass function to the IMF by a similar factor, $0.3$--$0.4$ \citep{AlvesLombardiLada2007,NutterWardThompson2007,EnochEtAl2008,Myers2008,AndreEtAl2010,KoenyvesEtAl2010,OffnerEtAl2014,FrankEtAl2014}. However, we warn that the claim that the core mass function can be mapped directly to the IMF, and that the observed core mass function is universal and has a robustly-detected turnover like the IMF, have both been subject to considerable dispute in the literature \citep{Dib10b,Krumholz14a,Bertelli-Motta16a,GuszejnovKrumholzHopkins2016,GuszejnovEtAl2018,LiptaiEtAl2017,Liu18a}; even if there is a link, the observed shift from the CMF to the IMF is not always $\approx 3$ \citep[e.g.,][]{Benedettini18a, Zhang18a}.

\paragraph{The Role of Magnetic Geometry}

Most previous simulations of magnetically-driven jet launching started from a uniform magnetic field aligned with the rotation axis of the core that forms the disk. However, in reality we expect a significant un-ordered, turbulent component to be present. That turbulent field component may either be inherited from the parent molecular cloud or be generated by small-scale dynamo processes \cite{BrandenburgSubramanian2005,SchekochihinEtAl2007,SurEtAl2010,FederrathSurSchleicherBanerjeeKlessen2011,FederrathEtAl2011,FederrathSchoberBovinoSchleicher2014,SchoberEtAl2012PRE,SchoberEtAl2015,SchleicherEtAl2013,Federrath2016jpp}.

\begin{figure*}
\centerline{\includegraphics[width=1.0\linewidth]{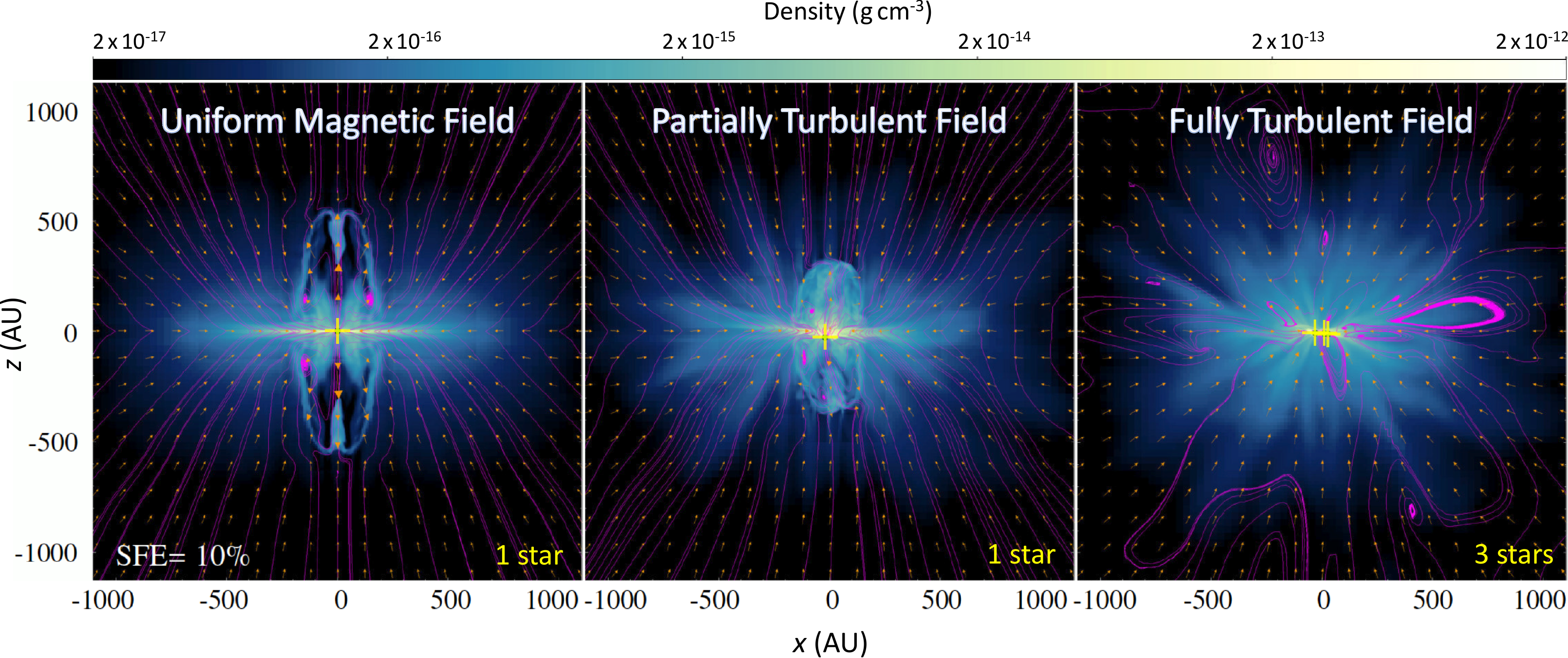}}
\caption{Protostellar disc and jet formation simulations with different magnetic field geometries. The left-hand panel shows the standard approach of using an initially uniform magnetic field aligned with the rotation axis of the core and disc. The middle panel adds a turbulent component to the uniform field component, such that both have the same rms. The right-hand panel show the same simulation, but with a completely turbulent magnetic field (no guide-field component present). Outflows are strongest in the uniform-field case, with a fast collimated jet component launched from the inner parts of the disk. Partially turbulent magnetic fields still generate an outflow, but weaker and less collimated. In the absence of a uniform field component, however, jets are completely suppressed, but fragmentation of the disk is induced, i.e., 3 stars form in the fully turbulent case, compared to only a single star in the other two simulations. Figure adopted from \citet{GerrardFederrathKuruwita2018}.}
\label{fig:g18}
\end{figure*}

\autoref{fig:g18} shows the results of recent simulations by \citet{GerrardFederrathKuruwita2018}, which the authors started with different magnetic field configurations in the core. They compare three simulations: one with a uniform field aligned with the rotation axis (left-hand panel), a second one that has a turbulent, tangled component in addition to the uniform field component, such that both have the same contribution to the total field strength (middle panel), and a third simulation that starts with a completely turbulent magnetic field without any ordered guide-field component (right-hand panel). The initial conditions and physics included in the three simulations are otherwise identical, and the total rms field strength is $100\,\mu\Gauss$ in all three cases.

We see in \autoref{fig:g18} that the uniform-field simulation produces a fast collimated jet aligned with the rotation axis of the disk. There is also a less-collimated wide-angle outflow component, but it does not carry much mass, compared to the case where both a uniform and turbulent magnetic field component is present (middle panel). This model is the most realistic and contains a fast collimated jet component and a wide-angle, low-speed outflow component previously seen in more idealized simulation setups \citep{BanerjeePudritz2006,MachidaEtAl2008,FederrathEtAl2014,KuruwitaFederrathIreland2017}, and consistent with recent ALMA observations in Serpens South \citep{HodappChini2018}. Both components may carry away a significant amount of mass. This is why in this model the protostar has the lowest accretion rate of all three cases, with a protostar mass of $0.15\,\msol$ after $1200\,\yr$, compared to $0.20\,\msol$ in the uniform-field case, at the same time after the protostar was formed.

A most striking result is the complete absence of an outflow in the fully turbulent field case. This demonstrates that an ordered magnetic field component aligned with the rotation axis of the disk is required to launch a magneto-centrifugally driven outflow, as described in the \citet{BlandfordPayne1982} mechanism of wind launching.

Overall, the accretion histories of the three simulations vary by up to 100\% -- for example, about $500\,\yr$ after protostar formation, the first protostar that forms in the fully turbulent field case has only reached $0.05\,\msol$, while the protostar in both the uniform-field and partially-turbulent field cases has a mass of about $0.10\,\msol$, i.e., significantly more massive, because the absence of addition magnetic-field pressure from the turbulent field component, which reduces the accretion rate onto the star. In summary, the magnetic field structure has significant impact on the jet launching and final mass of the protostar.

\subsubsection{Radiation Feedback and Magnetic Fields}
\label{ssec:mag_rad}

A number of authors have suggested the key physical process responsible for setting the location of the peak of the IMF is radiation feedback \citep{KrumholzKleinMcKee2007,Bate09a,Krumholz2011,GuszejnovKrumholzHopkins2016}. The central argument behind this hypothesis is that isothermal MHD turbulence is a scale-free process, and thus is incapable of producing a mass function with a characteristic scale such as the IMF. Consistent with this claim, simulations have shown that isothermal turbulence without feedback tends to produce fragmentation to arbitrarily small mass scales, leading to a mass function that is a pure power law, or that has a peak dependent on the resolution of the simulation, rather than a function with a distinct peak such as the observed IMF (\citealt{Bertelli-Motta16a,LiptaiEtAl2017,FederrathKrumholzHopkins2017,GuszejnovEtAl2018}; however, see \citealt{HaugboelleEtAl2018} for a contrasting view). On the other hand, radiative heating of a collapsing cloud by the protostars forming within it, whose luminosity is primarily powered by accretion, naturally does produce a characteristic mass scale that appears consistent with the observed IMF peak. Simulations that include radiation feedback generally yield IMFs that converge with resolution and are in reasonable agreement with observations \citep{Bate09a,Bate2012,Bate14a,OffnerEtAl2009,Krumholz11c,Krumholz12b,Myers14a,FederrathKrumholzHopkins2017,Cunningham18a,Li18a}.

In the context of such models, what is the role of magnetic fields? Simulations offer limited guidance, because most published work on the IMF including radiative transfer has either omitted magnetic fields entirely \citep{Bate09a,Bate2012,Bate14a,OffnerEtAl2009,Krumholz11c,Krumholz12b} or included it in all runs carried out \citep{Li18a}. The only published works on the IMF that perform a controlled experiment by including radiation feedback and repeating a calculation both including and omitting magnetic fields are those of \citet{PriceBate2009}, \citet{Myers14a}, and \citet{Cunningham18a}, and only the latter two of these also include outflows.\footnote{In the non-magnetized simulations the outflows are launched artificially via a sub-grid model, but this is also true in the magnetized simulations, since they do not have the resolution to follow outflow launching self-consistently while also running for long enough to allow meaningful statistical study of the IMF.} The general result of these studies is that, with the exception of their role in driving outflows, magnetic fields have only marginal effects on the final IMF. \citet{KrumholzEtAl2016} investigate why this should be by carrying out a detailed analysis of the simulations of \citet{Myers14a}; they show that, on the small scales ($\sim\mbox{few}\times 10^3$ AU) where protostellar cores fragment, thermal pressure support (enhanced by radiative heating) is generally stronger than magnetic support, even in simulations that are only marginally magnetically supercritical on large scales. The fundamental reason is that the processes that lead to the production of protostars involve gathering mass along field lines and possibly also turbulent reconnection \citep{Lazarian99a,Santos-Lima12a}, so that the dense regions near protostars that might or might not fragment, thereby determining stars' characteristic masses, have $\mu_\Phi$ values much larger than the average of the larger-scale cloud in which they are embedded.

We conclude this section by turning to the question of whether magnetic fields might play a larger role in shaping either the very low mass or very high mass parts of the IMF. On the massive end, the main distinguishing feature is that radiation feedback of massive stars is much more intense than that of low-mass stars, because for stars larger than $\sim 5$ $M_\odot$ Kelvin-Helmholtz contraction (supplemented by the onset of nuclear burning in stars larger than a few tens of $M_\odot$) produces a luminosity that rises sharply with mass. Simulations of the formation of such stars beginning from massive protostellar cores show that magnetic fields tend to aid in the growth of such stars via four mechanisms \citep{CommerconEtAl2010,CommerconEtAl2011,Myers13a}. First, they suppress fragmentation directly by providing magnetic support. Second, by providing a means of angular momentum transport, magnetic fields tend to make the disks of massive stars smaller, keeping the mass closer to the central star where it is warmer and less prone to fragment. Third, the enhanced angular momentum transport increases the accretion rate onto the central star, making it more massive and thus more luminous. Fourth and finally, by creating protostellar outflows, magnetic fields provide a ``vent'' that stops radiation from building up to the points where radiation pressure begins to inhibit accretion \citep{Krumholz05b,CunninghamEtAl2011,PetersEtAl2011,Kuiper15a,Kuiper16a}. While these effects have all been demonstrated in idealized simulations starting from initial massive cores, it is unclear whether they are significant for production of the IMF overall.

Radiation and magnetic fields interact in a different way for very low mass stars and brown dwarfs. The majority of such objects likely form by direct fragmentation in much the same manner as stars near the peak of the IMF \citep[e.g., see the review by][]{Chabrier14a}. However, formation via gravitational instability in the disk of a Solar-mass star represents a second possible formation channel, one for which we have direct observational evidence in at least some instances \citep{Tobin16a}. Magnetic fields (and non-ideal MHD effects) play a potentially-important role in modulating this channel, because they shape the properties of disks. In the extreme case of a protostellar core whose rotation axis is aligned with an initially-uniform magnetic field, and neglecting non-ideal effects, efficient magnetic braking prevents the formation of disks entirely \citep[e.g.,][]{Mellon08a, Hennebelle09b}, and thus necessarily prevents the formation of brown dwarfs or other low mass objects via disk instability. In reality magnetic fields certainly do not suppress disk formation entirely; Keplerian disks are observed even around the youngest protostars \citep[e.g.,][]{Tobin12a}. There are numerous candidate explanations for why disks persist, including misalignment of the rotation axis and the magnetic field \citep{Joos12a, Krumholz13a, Tsukamoto18a}, suppressed magnetic braking due to turbulence \citep{Seifried12a, Seifried13a}, and various non-ideal effects \citep{Santos-Lima12a, Tsukamoto15a, Tsukamoto18a}. Nonetheless, magnetic fields may reduce disk sizes compared to the purely hydrodynamic case, and smaller disks are in general more stable against self-gravity, because the matter is confined to regions where there is more stabilization by both shear and radiative heating from the central star \citep[e.g.,][]{Kratter10a}. Thus magnetic fields likely reduce the incidence of disk fragmentation \citep{BuerzleEtAl2011} and thereby suppress the disk formation channel for brown dwarfs and low mass stars. The amount of suppression is not yet known, since in the simulations carried out to date disk properties depend strongly on the assumed initial conditions. Moreover, even if magnetic fields do suppress disk fragmentation, it is not clear if this matters much for the overall IMF. Radiative heating by the central star renders disk fragmentation rare for stars near the peak of the IMF even in purely hydrodynamic simulations \citep[e.g.,][]{Bate09a, Bate2012, OffnerEtAl2009, Offner10a, Kratter10b}. Thus magnetic fields may simply further reduce a channel of brown dwarf formation that is already sub-dominant thanks to radiation feedback.

\subsection{Prospects and Future Work on the IMF}

We conclude that magnetic fields and feedback in the form of jets/outflows and radiation are important ingredients for understanding the IMF. Concerning magnetic fields in particular, there are two competing effects. On one hand, the magnetic field tends to directly reduce the fragmentation of cores and disks due to magnetic pressure and tension, therefore changing the physical conditions of the core and disk, even before stellar feedback starts. The importance of this effect in simulations appears to depend on whether the simulations also include other mechanisms that suppress fragmentation, particularly radiation. In simulations including radiation, increasing the strength of magnetic fields from zero up to a level where the star-forming cloud is only barely supercritical increases the median stellar mass by a factor of $\approx 1.5 - 2$. On the other hand, magnetic fields also drive powerful jets and outflows, which limit the stellar mass and induce fragmentation. This effect produces an effective core-to-star efficiency of about 1/3. This effective core-to-star efficiency is the result of two effects. First, each individual core loses about 1/2 of its mass in the individual outflow of that core. Second, the outflows induce additional fragmentation of the filaments that feed the cores, thereby reducing the average star mass further by another factor of $\sim2/3$. Because this feedback effect is comparable in magnitude but opposite in direction to the effects of magnetic fragmentation suppression, the net effect of both processes is to alter the location of the IMF peak at the factor of $\approx 2$ level, smaller than one might expect based on consideration of either process alone.

Not only the magnitude of the magnetic field, but also its structure (ordered versus turbulent) plays a critical role in controlling the strength of the outflows and in determining the resulting mass distribution of stars. Recent observational studies, for example with ALMA, are now beginning to reveal the complex magnetic field structures inside cores and disks, and in the outflows \citep{HullEtAl2017ALMA,ZhangEtAl2018,CoxEtAl2018}, often showing turbulence and significant deviations from the classic hourglass shape. More observational constraints on the magnetic field geometry are needed to inform theoretical models and simulations.

In addition to the challenges in understanding the IMF at present day, we need to work even harder to understand what the mass function of primordial stars might have been. Observations so far can only provide indirect constraints on the mass of the first stars in the Universe. Simulations of the formation of the first stars indicate that the disks in which they form can fragment even under the conditions of primordial chemistry and cooling \citep{ClarkEtAl2011,GreifEtAl2011,SusaEtAl2014,HiranoEtAl2015}. However, an important limitation of these studies is that they neither include magnetic fields nor jet/outflow feedback, both of which may play a crucial role also in primordial star formation \citep{Federrath2018,Klessen19a}.

\section{Summary and Prospects}

Our view of the importance of magnetic fields in the process of star formation -- and particularly in determining the two most important outputs of that process, the overall star formation rate (SFR) and the initial mass function (IMF) of stars -- has changed dramatically over the last 15 years. Prior to that time, most theoretical models of star formation assigned magnetic support a key role in setting both the the SFR and the IMF. With the discovery that magnetic fields in star-forming regions are not as strong as once believed, so that most star-forming regions are magnetically supercritical, this view has shifted. In a supercritical region, magnetic fields cannot directly inhibit collapse by a substantial amount, and so magnetic fields alone cannot provide an explanation for the surprisingly-low rate of star formation that we observe on all scales, from individual clouds near the Sun to entire galaxies. Nor can they by themselves regulate the fragmentation of collapsing gas and thereby provide an explanation for the apparently universal or near-universal mass scale of stars.

While magnetic fields are no longer the star of the show, modern theoretical models and simulations tuned to match observed field strengths indicate that they still play a non-negligible supporting role. By providing resistance to turbulent compression and pressure that opposes gravity, magnetic fields directly reduce the ability of turbulence to gather gas into gravitationally-unstable clumps. This lowers the star formation rate by a factor of $2-3$ compared to the outcome in non-magnetized flows, and increases the median mass of those clumps that do become unstable and go on to form stars by a similar factor. The strength of this effect can be measured in simulations, but is not completely understood analytically, as it depends critically on how magnetic field strengths vary with density in a medium where the turbulence is supersonic and trans-Alfv\'enic. While we have a reasonable model for this correlation in super-Alfv\'enic flows, our model breaks down in the trans-Aflv\'enic regime that is more likely to characterize star formation. Progress toward a quantitative  analytic understanding of how magnetic fields reduce the star formation rate and raise the mean mass of star-forming regions will require an extension of our understanding of the magnetic field-density correlation to this regime.

Magnetic fields also play a critical indirect role by providing the means for forming stars to launch jets and outflows. On small scales, outflows lower the mean stellar mass by a factor of $\approx 3$, through a combination of ejecting gas that would otherwise accrete onto stars and by encouraging fragmentation. This nearly counters the effects of magnetic support in shifting the IMF to higher values, so that the combined effects of magnetic suppression of fragmentation and outflows is to change the mean stellar mass by only a factor of $\approx 2$ compared to the outcome in a non-magnetized flow. On larger scales, outflows help stir turbulent motions in clouds and directly eject mass from collapsing regions, thereby lowering the rate of star formation by an additional factor of several compared to magnetized clouds without outflows, and by an order of magnitude or more compared to the case of purely hydrodynamic turbulence. Magnetic fields may also slow the rate of turbulent decay in collapsing clouds outright, although this prospect has thus far been demonstrated only in idealized compressing box simulations.

While the interplay of magnetic fields and outflows is now reasonably well if not fully understood, the interaction of magnetic fields with other forms of stellar feedback has been explored far less extensively. If there is any possibility for magnetic fields to return to a starring role in models of star formation, it lies in these unexplored frontiers. We highlight one particularly interesting prospect for further investigation, which is that magnetic fields might fundamentally change the way that hot gas interacts with the cold ISM, by reducing the rate of material and thermal exchange across hot-cold gas interfaces. This could potentially make supernovae feedback much more effective than currently suspected, which in turn would have major implications for the star formation rate and, on larger scales, for the properties of galactic winds. However, this is just one example -- the interaction of magnetic fields with other types of feedback is equally-poorly known, and the possibility remains that magnetic effects will again prove crucial to models of star formation.

\section*{Conflict of Interest Statement}

The authors declare that the research was conducted in the absence of any commercial or financial relationships that could be construed as a potential conflict of interest.

\section*{Author Contributions}

MRK and CF jointly wrote all parts of this review.

\section*{Funding}
MRK and CF acknowledge support from the Australian Research Council's \textit{Discovery Projects} and \textit{Future Fellowship} funding schemes, award numbers DP150104329 (CF), DP160100695 (MRK), DP170100603 (CF), FT180100375 (MRK) and FT180100495 (CF), and the Australia-Germany Joint Research Cooperation Scheme (UA-DAAD; both MRK and CF).

\section*{Acknowledgments}
MRK and CF both acknowledge the assistance of resources and services from the National Computational Infrastructure (NCI), which is supported by the Australian Government. We further acknowledge high-performance computing resources provided by the Leibniz Rechenzentrum and the Gauss Centre for Supercomputing (grants~pr32lo, pr48pi and GCS Large-scale project~10391; CF), the Partnership for Advanced Computing in Europe (PRACE grant pr89mu; CF), and Australian National Computational Infrastructure grants jh2 (MRK) and ek9 (CF) in the framework of the National Computational Merit Allocation Scheme and the ANU Allocation Scheme. We thank A.~Tristis, B.~Gerrard, and two anonymous referees for helpful comments.



\bibliographystyle{frontiersinSCNS_ENG_HUMS} 
\bibliography{allrefs}





\end{document}